\documentclass[aps,twocolumn,10pt,superscriptaddress,prl]{revtex4-1}

\usepackage{amsmath,amssymb,amsfonts}
\usepackage{graphicx}
\usepackage{braket}
\usepackage{nicefrac}
\usepackage{multirow}
\usepackage{cellspace}
\usepackage[colorlinks=True,linkcolor=red,citecolor=blue,urlcolor=blue]{hyperref}
\usepackage[retainorgcmds]{IEEEtrantools}
\usepackage{xcolor}
\usepackage[title]{appendix}

\begin{document}

\title{Quantization in Chiral Higher Order Topological Insulators: \\Circular Dichroism and Local Chern Marker}

\author{Oscar Pozo}
\affiliation{Instituto de Ciencia de Materiales de Madrid,
and CSIC, Cantoblanco, 28049 Madrid, Spain}
\author{C\'{e}cile Repellin}
\affiliation{Department of Physics, Massachusetts Institute of Technology, Cambridge, Massachusetts 02139, USA}
\author{Adolfo G. Grushin}
\affiliation{University Grenoble Alpes, CNRS, Grenoble INP, Institut N\'eel, 38000 Grenoble, France}

\begin{abstract}
The robust quantization of observables in units of universal constants is a hallmark of topological phases. We show that chiral higher order topological insulators (HOTIs), bulk insulators with chiral hinge states, present two unusual features related to quantization. First, we show that circular dichroism is quantized to an integer or zero depending on the orientation of the sample. This probe locates the hinge states, and can be used to distinguish different types of chiral HOTIs. Second, we find that the average of the local Chern marker over a single surface, an observable related to the surface Hall conductivity known to be quantized in the infinite slab geometry, is nonuniversal for a finite surface. This is due to a nonuniversal contribution of the hinge states, previously unaccounted for, that distinguishes surfaces of chiral HOTIs from Chern insulators. Our findings are relevant to establish higher order topology in systems such as the axion insulator candidate EuIn$_2$As$_2$, and cold atomic realizations.
\end{abstract}

\maketitle

\textit{Introduction.}---The quantization of global observables is a quantum mechanical signature that distinguishes trivial and topological matter. For example, two-dimensional (2D) insulators without time-reversal symmetry present a Hall conductance that is quantized to $C e^2/h$ where $C$ is an integer known as the Chern number. In contrast, experiments that show a quantized observable in three dimensions (3D) remain scarce; the sole examples are the observation of a quantized optical rotation in 3D time-reversal invariant topological insulators~\cite{wu2016quantized,Dziom:2017uh},
%\cite{wu2016quantized,Okada:2016tk,Dziom:2017uh}
and the quantization of a photocurrent in chiral topological semimetals~\cite{Rees:2019ue}. 

Our first result is that 3D chiral higher order topological insulators (HOTIs) have a quantized circular dichroism. HOTIs~\cite{Benalcazar,Benalcazar:2017cn,schindler2018higher,Song:2017eva,Sitte:2012ib,Zhang2013,PhysRevLett.119.246401,Wakabayashi2017,Ezawa:2018gt,ezawa2018magnetic,Ahn:uc,Wang2018,Wieder:2018qpe,Lin2017,khalaf2018higher,varnava2018surfaces,Kooi:2018cr,vanMiert2018,Franca2018,Trifunovic2019,Matsugatani:2018jb,Song:2018cj,Schindler2018bism, SessiScience2016} are a subclass of topological insulators~\cite{HasanKane2010,Qi2011} for which a $d$-dimensional bulk topological invariant protects metallic boundary modes in dimension $d-2$ or lower. More specifically, 3D chiral HOTIs~\cite{Sitte:2012ib,Zhang2013,schindler2018higher,PhysRevLett.119.246401,ezawa2018magnetic,Ahn:uc,Wieder:2018qpe,khalaf2018higher,Kooi:2018cr,varnava2018surfaces,vanMiert2018,Matsugatani:2018jb} are fully gapped with only conducting chiral hinge states. The protection is guaranteed by combinations of crystalline and discrete symmetries, but symmetry alone does not necessarily dictate which hinges have a boundary mode. 

\begin{figure}[t]
    \centering
    \includegraphics[scale=1.0]{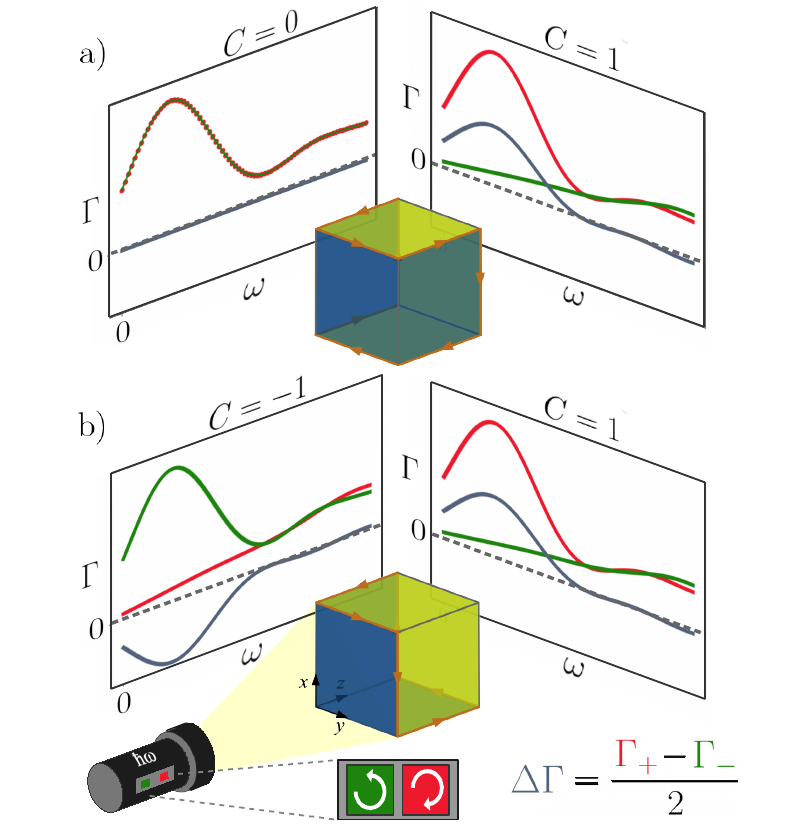}
    \caption{The frequency integrated in-gap absorption difference between left- and right-circularly polarized light (red and green lines, respectively) is quantized to $C$ depending on the polarization plane.  The cubes depict chiral HOTIs with hinge modes protected by (a) $C_{4}^{z}\mathcal{I}$- or (b) $\mathcal{I}$ symmetries. The light green and dark blue surfaces signal positive and negative values of the local Chern marker. Orange hinges represent gapless chiral states.}
    \label{fig:SchematicCD}
\end{figure}

Circular dichroism is an optical probe that is sensitive to topological properties~\cite{WangHsieh2011,Goldman2012,Souza2008,WangGedik2013,Schuler2017,de_juan_quantized_2017,LiuYing2018,Tran2017,TranCooper2018,Repellin2019,Schuler2019}.
It is the differential absorption of left and right circularly polarized light and its frequency integral is determined by the dc Hall conductivity through a sum rule~\cite{Bennett:1965jy,Souza2008}.
For 2D Hall insulators this integral is quantized to the Chern number for periodic boundary conditions~\cite{Tran2017,Repellin2019}, and vanishes upon including the edge state contribution~\cite{TranCooper2018,Tran2017,Souza2008}. To filter out the edge contribution, it is possible to use optical selection rules in Landau levels~\cite{TranCooper2018}, or, in periodically driven cold-atomic gases, a trap-release protocol~\cite{Tran2017} or a harmonic trap as was demonstrated experimentally~\cite{Asteria:2019if}.
In 3D we show that the circular dichroic absorption of chiral HOTIs restricted to a frequency range within the energy gap but {\it across the whole sample} is determined by an integer. The integer depends on the orientation of the sample and thus may be used to distinguish different types of chiral HOTIs (see Fig.~\ref{fig:SchematicCD}).

In addition to bulk properties, the topological character of topological insulators can emerge as a quantized surface property. The surface Hall conductivity of a chiral HOTI is quantized in units of $e^2/2h$, half of the conductance quantum~\cite{varnava2018surfaces,Coh:2011gq}. This quantization is inherited from the Hall conductivity of a gapped topological insulator surface state~\cite{Essin:2009kb,QiHughesZhang08} and can be derived in the infinite slab geometry~\cite{varnava2018surfaces,rauch2018geometric}; each surface Dirac cone contributes with $\pm e^2/2h$ to the total surface Hall conductivity. 

Our second result revises such quantization of surface observables; we show that for finite samples, the hinge states modify the surface quantization of the Hall conductivity by a nonuniversal factor. We use a real space version of the Berry curvature, known as the local Chern marker~\cite{bianco2011mapping}. It measures the local (space resolved) contributions to the Hall conductivity. In 2D insulators, the bulk average of the local Chern marker is quantized and equal to the Chern number $C$ of the corresponding periodic system. For open boundary conditions, its average over the whole sample vanishes. It means that the total contribution from edge states is equal and opposite in sign to the bulk contribution~\cite{bianco2011mapping,TranCooper2018,Tran2017}. We find that this does not happen for a finite surface of chiral HOTIs due to the spatial distribution of the local Chern marker in the hinge states, which distinguishes surfaces of HOTIs from purely 2D Chern insulators, despite their spectral similarities~\cite{Matsugatani:2018jb}. 

\textit{Models.}---We consider a generic tight-binding model of chiral HOTIs on the cubic lattice~\cite{schindler2018higher,khalaf2018higher,ezawa2018magnetic}
\begin{IEEEeqnarray}{rCl}
\label{eq:model}
   \hat{H}_{0} & = & \Big( M+J\sum_{i=1}^{3}\cos(k_{i}a) \Big) \tau_{3}\sigma_{0} + \sum_{i=1}^{3}\lambda_i\sin(k_{i}a) \ \tau_{1}\sigma_{i} + \nonumber \\
   & + & D\big( \cos(k_{1}a) - \cos(k_{2}a) \big) \tau_{2}\sigma_{0} + \tau_{0}\mathbf{B}\cdot\boldsymbol{\sigma}  \  ,
\end{IEEEeqnarray}
where $k_{i}$ are the cartesian momentum components, $a$ is the lattice spacing, $\sigma_{i}, \tau_{i}$ are two sets of Pauli matrices acting on the spin and orbital degrees of freedom and $\sigma_0,\tau_0$ are $2\times 2$ identity matrices. At half-filling the first two terms describe a time-reversal invariant 3D topological insulator, with anisotropic Fermi velocities if $\lambda_i$ are chosen to be different. The third term, proportional to $D$, breaks $C_{4}^{z}$ and time-reversal ($\mathcal{T}$) symmetries, but preserves their combination. The last term proportional to $\mathbf{B}$ is a Zeeman term that breaks $\mathcal{T}$ but preserves inversion symmetry ($\mathcal{I}$). The Hamiltonian Eq.~\eqref{eq:model} interpolates between a topological insulator ($D=|\mathbf{B}|=0$), 
a chiral HOTI protected by the rotoinversion symmetry $C_{4}^{z}\mathcal{I}$ [Fig. \ref{fig:SchematicCD} (a)] and a chiral HOTI protected by $\mathcal{I}$ [Fig. \ref{fig:SchematicCD} (b)]. We choose the parameters that maximize the bulk gap and therefore the localization of hinge states. Specifically, $M/J=2$, $\lambda_i/J = 1$ (all three velocities are the same unless otherwise stated) and  $|\mathbf{B}|/J=0.5$, with $D/J=0$ ($D/J=1$) and $\mathbf{B} \parallel (1,1,1)$ ($\parallel (0,0,1)$), for $\mathcal{I}$- ($C^{z}_4\mathcal{I}$-) symmetric HOTIs (see also \cite{SupplementalHOTI} sec. A).
\\
\begin{figure}
    \centering
    \includegraphics[scale=1.0]{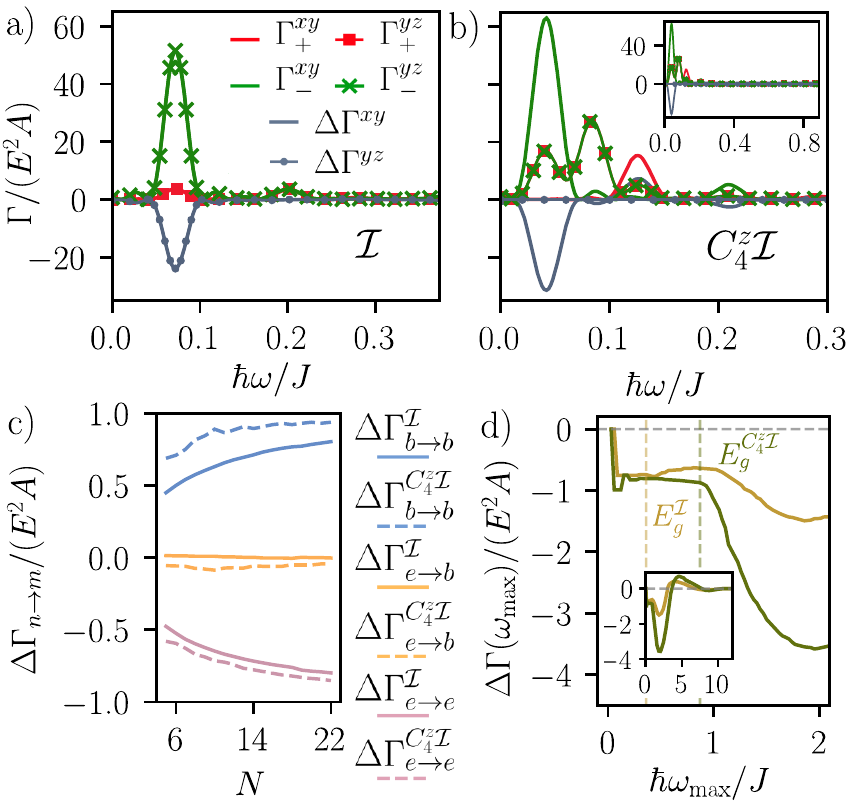}
    \caption{Distinguishing $C^{z}_4\mathcal{I}$- and $\mathcal{I}$-symmetric HOTIs with circular dichroism. (a), (b) In-gap absorption of the circular signal at frequency $\omega$ for left (red) or right (green) polarization or the difference between the two (blue), for polarization planes $xy$ and $yz$. We choose $N_{\mathrm{tot}}= N^{3}$ sites, with $N=17$, $t J/\hbar=200$. In the $\mathcal{I}$ HOTI, the integral of the blue curve in (a) is $C = -0.73, -0.73$ for $xy$ and $yz$ polarization, respectively. In contrast, in the $C^{z}_4\mathcal{I}$ HOTI, (b), we find $C = -0.86, 0$. (c) Contributions from edge (e) and bulk (b) states as a function of system size, illustrating the convergence to a quantized (or vanishing) value. (d) Integrated dichroic signal, Eq.~\eqref{integrated dichro}, as a function of integration cutoff. The position of the gap $E_g$ is indicated by a vertical line. The absorption $\Gamma$ was obtained using Eq.~\eqref{eq:CDEnergy} [(a),(b)] or its approximation in the long-time limit [(c),(d)]. See \cite{SupplementalHOTI} sec. B for more details.}
    \label{fig:CD}
\end{figure}
\textit{Quantized circular dichroism of chiral HOTIs.}---To first order in perturbation theory, the absorption rate $\Gamma(\omega)$ of a circular periodic excitation of intensity $E^2$ polarized in the $xy$ plane at frequency $\omega$ is  (see \cite{SupplementalHOTI} sec. B for details)
\begin{equation}
    \dfrac{\Gamma_{\pm}(\omega)}{2\pi E^{2}} =
    \sum_{n,m} \big| \bra{m}\left( \hat{x}\pm i\hat{y} \right) \ket{n} \big|^{2} \delta^{(t)}(E_{m}-E_{n}-\hbar\omega),
    \label{eq:CDEnergy}
\end{equation}
where $\ket{n}$ ($\ket{m}$) are occupied (unoccupied) states with energy $E_n$ ($E_m$) and $\delta^{(t)}(\epsilon) \equiv (2\hbar /\pi t) \sin^2(\epsilon t/2\hbar)/\epsilon^2$ (analogous definitions hold for other cartesian planes). We assume half-filling and discuss changing the chemical potential in~\cite{SupplementalHOTI} sec B.
In the long time limit $\delta^{(t)}(\epsilon)$ is the Dirac delta function, so that Eq.~\eqref{eq:CDEnergy} becomes the familiar Fermi's Golden Rule. The integrated differential rate of absorption is defined as
% % 
\begin{equation}
    \Delta\Gamma(\omega_\mathrm{max}) \equiv \int_{0}^{\omega_\mathrm{max}} d\omega  \left[ \Gamma_{+}(\omega)-\Gamma_{-}(\omega) \right] /2.
    \label{integrated dichro}
\end{equation}
In the limit where $\omega_\mathrm{max}\to \infty$ it simplifies to~\cite{Souza2008,Tran2017}
\begin{equation}
    \Delta\Gamma(\infty) = -2\pi i E^{2} \ \text{Tr}\Big\{ \big[ \hat{Q}\hat{x},\hat{P}\hat{y} \big] \Big\} \equiv E^2 \sum_{\text{all } \textbf{r}_{i}}C_{xy}(\textbf{r}_{i}).
    \label{CDLCM}
\end{equation}
This equality---valid in two and three dimensions---expresses the dichroic signal as the sum of the local Chern marker $C_{xy}(\textbf{r}_{i})$ defined as a trace of the complete set of Wannier states $\left\{ \ket{\textbf{r}_{i}} \right\}$.
$\hat{P}$ ($\hat{Q}$) projects onto the occupied (unoccupied) eigenstates, and $\hat{x}$, $\hat{y}$ are position operators. In a 2D insulator with periodic boundary conditions the average of the local Chern marker is equal to the Chern number~\cite{bianco2011mapping,tran2015topological}. Generically, for a finite system with open boundary conditions the average of $C_{xy}(\textbf{r}_{i})$ over the whole sample vanishes due to the trivial character of the fiber bundle~\cite{bianco2011mapping}, leading to $\Delta\Gamma(\infty)=0$. To show this mathematically recall that the trace of the commutator in Eq.~\eqref{CDLCM} vanishes in a finite vector space.

Our main result is that the integrated in-gap circular dichroism is finite and quantized; by choosing the frequency cutoff in Eq.~\eqref{integrated dichro} to be $\hbar\omega_\mathrm{max} = E_g$, where $E_g$ is the gap of the infinite slab geometry, $\Delta\Gamma(E_g)$ isolates the quantized hinge state contribution.

To reach this result, consider first periodic boundary conditions in the polarization plane, in which case only bulk to bulk transitions contribute and $\Delta\Gamma(\infty) =\Delta\Gamma_{b\rightarrow b}$. 
From Eq.~\eqref{CDLCM}  $\Delta\Gamma_{b\rightarrow b}$ is proportional to the integral of the Chern marker, which is not an integer in general~\cite{Tran2017}. 
For a 2D insulator with area $A$, $\Delta\Gamma_{b\rightarrow b}=CE^{2}A$ where $C$ is the Chern number~\cite{Tran2017}. For a 3D chiral HOTI slab with surface area $A$ only the two surfaces parallel to the polarization plane contribute to  $\Delta\Gamma_{b\rightarrow b}$; their surface Dirac cones contribute with $\pm 1/2$ each, resulting in a total of
\begin{equation}
\label{eq:bulkbulk}
    \Delta\Gamma_{b\rightarrow b} = CE^{2}A  \  ,
\end{equation}
where $C$ is the sum of these contributions, and thus an integer~\cite{varnava2018surfaces}.
We have checked this expectation numerically (see~\cite{SupplementalHOTI} sec. B).

In the more realistic case of open boundary conditions there are four types of $n \to m$ transitions contributing to Eq.~\eqref{CDLCM}: from bulk to bulk ($b \to b$), bulk to edge ($b \to e$), edge to bulk ($e \to b$) and edge to edge ($e \to e$) states. The sum of all transitions must satisfy
\begin{eqnarray}
\label{eq:edgeedgeCD}
    \Delta\Gamma(\infty)=\sum_{n,m \in \lbrace e,b\rbrace}\Delta\Gamma_{n\rightarrow m} = 0 \ .
\end{eqnarray}
We have evaluated the relative sizes of the different contributions numerically using Eq.~\eqref{eq:CDEnergy} in the long time limit for the $\mathcal{I}$-
and $C_4\mathcal{I}$-symmetric HOTIs [see Fig. \ref{fig:CD} (c)]. In all cases the edge-bulk and bulk-edge transitions become negligible as we increase the system size; this is related to the exponentially small overlap between the surface and bulk states (see \cite{SupplementalHOTI} sec. B). In virtue of Eq.~\eqref{eq:edgeedgeCD} the bulk-bulk and edge-edge transitions are equal and opposite, and due to Eq.~\eqref{eq:bulkbulk} their respective thermodynamic values are $\pm C E^2A$. Thus, restricting the integration of the circular dichroic signal to in-gap absorption singles out edge-edge transitions, which guarantees the quantization of $\Delta\Gamma(E_g)$.

In Figs.~\ref{fig:CD} (a) and (b) we show the circular dichroism calculated by finite time evolution in the frequency range $\omega\in(0,E_g)$ for $\mathcal{I}$- and $C_{4}^{z}\mathcal{I}$-symmetric HOTIs. By evaluating $\Delta\Gamma(E_g)$ for $N_\mathrm{tot}=N^3$ sites with $N=17$ for $xy$ polarization we obtain $C = -0.73$ and $-0.86$ for the $\mathcal{I}$- and $C_{4}^{z}\mathcal{I}$-symmetric HOTIs, respectively. As $N$ grows, $C$ approaches exact quantization as seen in Fig.~\ref{fig:CD} (c).
For $yz$ polarization, we obtain $C = -0.73$ and $0$ for the $\mathcal{I}$- and $C_{4}^{z}\mathcal{I}$-symmetric HOTIs, respectively.

\begin{figure*}
    \centering
    \includegraphics[scale=1]{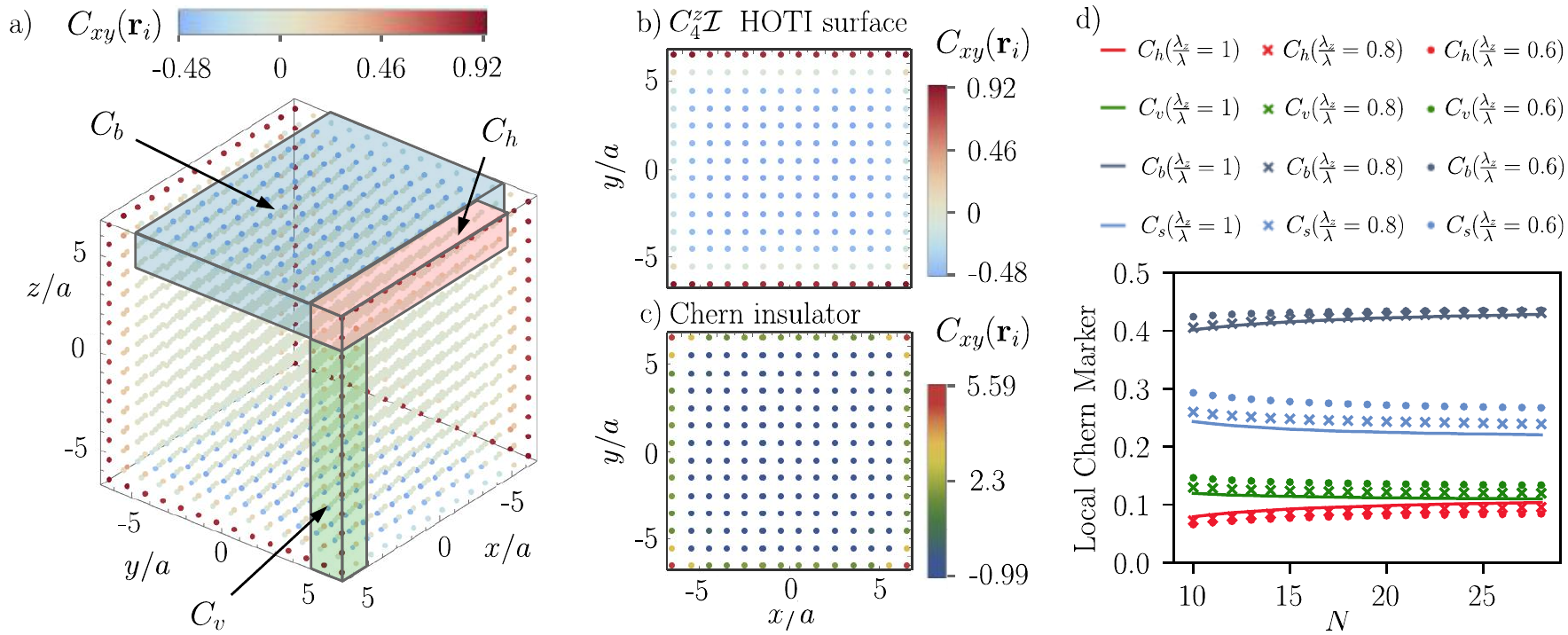}
    \caption{(a) Local Chern marker $C_{xy}(\textbf{r}_{i})$ of the $C_4^{z}\mathcal{I}$-symmetric HOTI. In the top surface (hinge excluded, highlighted in blue) it is quantized to $-1/2$. The enlargement of this surface cut (b) shows the difference with the local Chern marker distribution in a Chern insulator (c). (d) Local Chern marker integrated over one vertical ($C_v$) or horizontal ($C_h$) hinge, or over the top surface hinge excluded ($C_b$) or included ($C_s = 2C_h + C_b$). The Fermi velocities of the HOTI model are either isotropic ($\lambda_z/\lambda = 1$) or anisotropic ($\lambda_z/\lambda = 0.6, 0.8$). For these calculations, we have considered surface and hinges with a two-site depth and normalized over the area of the blue region. The data in d) have been calculated using the \texttt{Kwant} package~\cite{Groth_2014}, with a broadening of $10^{-3}$ (see also \cite{SupplementalHOTI} sec. C).
    }
    \label{fig:tetrahedral}
\end{figure*}

An experimentally useful property of circular dichroism is that it depends on the type of chiral HOTI and the polarization plane. This follows from Fig.~\ref{fig:CD} and is depicted schematically in Fig.~\ref{fig:SchematicCD}. For the $\mathcal{I}$-symmetric HOTI, the circular dichroic signal is always nonzero, and quantized in the thermodynamic limit. However, for the $C^z_4\mathcal{I}$-symmetric HOTI, $C$ vanishes for $xz$ and $yz$ polarized light, but it remains finite for $xy$ polarization. 
Indeed, in the $C^z_4\mathcal{I}$ HOTI the projection of the hinge states onto the $xy$ plane results in an orientable closed path akin to the edge state of a Chern insulator.
This path contributes an integer $C\neq 0$ to the circular dichroic signal, in contrast to the $xz$ and $yz$ projections, where no such path exists. On the contrary, in the $\mathcal{I}$-symmetric HOTI, all cartesian polarizations  lead to orientable
closed paths, and the circular dichroic signal is quantized. 
In general, the quantization is determined by the projection of the 3D hinge state onto the polarization plane, which might be non-Cartesian. In this plane, each orientable closed path contributes an integer $C\neq 0$ to the circular dichroic signal 
normalized by its area, and with a sign dictated by the orientation of the path. 

Our characterization scheme relies on the in-gap ($\omega < E_g$) integration of the dichroic signal.
 Although \emph{ab initio} methods may be able to approximate $E_g$, it is desirable to determine it experimentally. This is possible by measuring $\Delta \Gamma(\omega_\mathrm{max})$ as a function of $\omega_{\mathrm{max}}$, as illustrated Fig.~\ref{fig:CD} (d).
We expect a singularity at $\hbar\omega_\mathrm{max}=E_g$, caused by the activation of bulk-to-bulk transitions. The additional singularities at $\hbar\omega_\mathrm{max}< E_g$ in Fig.~\ref{fig:CD} (d) are due to the discretization of hinge states in finite size, and will not affect the extraction of the gap in the thermodynamic limit.

\textit{Local Chern marker of chiral HOTIs.}---So far we have taken advantage of the energy resolution of circular dichroism to predict a quantized dichroic signal. Can sacrificing energy resolution in favor of a local measurement lead to a universal quantized observable? In infinite slabs the answer is affirmative: the Hall conductivity of the surface of a chiral HOTI is quantized in units of $\pm e^2/2h$~\cite{Essin:2009kb,Mong:2010ka,Coh:2011gq}. 
It stems from the quantization to $\pm1/2$~\cite{varnava2018surfaces,rauch2018geometric} of the local Chern marker $C_{xy}(\textbf{r}_{i})$~\cite{bianco2011mapping}, the dimensionless operator defined in Eq.~\eqref{CDLCM}, see also \cite{SupplementalHOTI} sec. B and C.
We show next that the situation is different for finite surfaces: the hinge contribution renders the surface average of the local Chern marker nonuniversal.
Our results carry over to other choices of position operators in Eq.~\eqref{CDLCM}.

We have calculated the local Chern marker using a recent approach~\cite{Varjas:2019wz} based on the kernel polynomial method~\cite{Weisse:2006goa} implemented using the \texttt{Kwant} package~\cite{Groth_2014}. It allows us to reach sizes beyond exact diagonalization, and up to $N_\mathrm{tot}=N^3$ sites with $N=28$. For concreteness, we focus on the distribution of $C_{xy}(\textbf{r}_{i})$ for the $C_4^{z}\mathcal{I}$-symmetric HOTI, shown in Fig.~\ref{fig:tetrahedral} (a). $C_{xy}(\textbf{r}_{i})$ is close to zero deep in the bulk of the cube. The largest contribution to $C_{xy}(\textbf{r}_{i})$ comes from the $xy$ surfaces and decays fast with increasing distance to the surface.

Consider the top $xy$ surface of the HOTI, as represented in Fig.~\ref{fig:tetrahedral} (b). Far from the hinges, we expect the local Chern marker to approach $-1/2$, since the Hall conductivity is quantized to $-e^2/2h$ in the infinite slab geometry~\cite{Essin:2009kb,Mong:2010ka,Coh:2011gq,varnava2018surfaces} (see also \cite{SupplementalHOTI} sec. B and C).
Our numerical results indeed show this convergence as a function of system size at the central points of the surface (see Fig.~\ref{fig:tetrahedral} a) and b). Remarkably, the average of the local Chern marker over a single surface of the HOTI does not vanish (see Fig.~\ref{fig:tetrahedral} d)). This is in stark contrast with 2D Chern insulators (see Fig.~\ref{fig:tetrahedral} c)) and highlights the 3D character of the HOTI.

The average of the local Chern marker over the surface of a HOTI is finite, but is it still universal? Note that the sum of all hinge contributions must exactly compensate the bulk, a fact we have numerically checked. 
However, $C^{z}_4\mathcal{I}$ symmetry only imposes that hinges related by this symmetry have equal weight. In Fig.~\ref{fig:tetrahedral} d) we show that the Chern marker can be distributed inhomogeneously over the horizontal and vertical hinges by choosing $\lambda_x=\lambda_y\equiv \lambda\neq\lambda_z$. These parameters introduce hopping anisotropy but do not break the $C_4^{z}\mathcal{I}$ symmetry that protects the topological state.  As $N$ increases, and up to numerical precision (see also \cite{SupplementalHOTI} sec. C), different parameters lead to a different distribution of the local Chern marker along the hinges. This means that although the total hinge contribution is fixed and equal to one, it can be unevenly distributed over the horizontal and vertical hinges. This in turn, implies that the surface average of the Chern marker is a non-universal number.

\textit{Discussion.} 
We have shown that the in-gap circular dichroism, a spatially averaged, frequency integrated quantity is quantized and can differentiate different types of chiral HOTIs. 
%%%%%%%%%%%%%%%%%%%%%%%
% Previous manuscript %
%%%%%%%%%%%%%%%%%%%%%%%
% It may be considered as an alternative to optical rotation measurements~\cite{wu2016quantized,Okada:2016tk,Dziom:2017uh}. Advantageously, the substrate does not affect circular dichroism, so long as it preserves time-reversal symmetry.
%%%%%%%%%%%%%%%%%%%%%%%
%\noteOP{Proposal:
%Although experimentalists currently measure rotations\noteCR{polarization rotations?}~\cite{wu2016quantized,Okada:2016tk,Dziom:2017uh}, we propose that the integrated circular dichroic response could be a superior way of measuring quantized optical responses; it does not depend on the substrate's details as long as it preserves time-reversal symmetry.}\noteAG{I iterated.}
%\noteAG{Why are there ?? in the refs?}
Although experimentalists currently measure polarization rotations~\cite{wu2016quantized,Dziom:2017uh}, we propose integrated circular dichroism as an advantageous alternative to measure quantized optical responses; importantly, it does not depend on the substrate's details as long as it preserves time-reversal symmetry.
Specifically, it could  probe higher-order topology in EuIn$_2$As$_2$~\cite{Xu:2019ui}, CrI$_3$/Bi$_2$Se$_3$/MnBi$_2$Se$_4$ heterostructures~\cite{Hou:2019wd}, and MnBi$_2$Te$_4$~\cite{Zhang:2019fk,Otrokov:2018vx,Gong:2018to}, all of which are predicted to host chiral hinge states. 
In particular, a quantized signal can be expected for a sample of EuIn$_2$As$_2$, illuminated with light polarized in a plane perpendicular to any of the crystal axis, and integrated up to the gap energy ($E_g\sim10m$eV from ab-initio calculations~\cite{Xu:2019ui}). These energies are comparable to those already used to measure quantized optical rotation~\cite{wu2016quantized}. 
Moreover, the quantization of circular dichroism applies to interacting systems~\cite{Repellin2019}, which suggests the possibility to distinguish different interacting HOTIs~\cite{Tiwari2019}. Lastly, quantized circular dichroism~\cite{Tran2017} has been measured in ultra-cold atomic Chern insulators \cite{Asteria:2019if} opening up the possibility to measure quantization in synthetic chiral HOTIs.

We have also found that the surface average of the local Chern marker, a spatially resolved quantity integrated over all energies, is non-universal in chiral HOTIs due to the hinge states, in sharp contrast to the infinite slab geometry and to a 2D Chern insulator.
As a consequence, an experiment aiming to characterize a chiral HOTI based on its surface properties must distinguish bulk and hinge contributions. 
In solid state systems the local Chern marker is related to the orbital magnetization~\cite{BiancoResta2013}. In cold atom experiments, it could be measured using a quantum gas microscope~\cite{Caio:2019jk,ardila2018measuring}. Mechanical~\cite{mitchell2018amorphous} and photonic topological metamaterials~\cite{schine2019electromagnetic} may offer additional avenues to measure this quantity.

We expect our work to motivate new experiments that can push the ongoing effort to measure quantization in 3D~\cite{wu2016quantized,Okada:2016tk,Dziom:2017uh,Rees:2019ue}, and topology in real space~\cite{Caio:2019jk,mitchell2018amorphous}. It calls to revisit 3D topological markers~\cite{Olsen:2017bz,Malashevich:2010hn,Taherinejad2015} to complete our understanding of topological matter in real space. 
\\

\textit{Acknowledgements}
We are thankful to B. A. Bernevig, D. Beno\^{i}t, F. Jean-Jacques, F. de Juan, T. Neupert, P. San Jose, I. Souza, D. Varjas, and M. A. H. Vozmediano for support and useful conversations. We are thankful to N. Goldman for his critical reading of the manuscript.
A. G. G. is supported by the ANR under the grant ANR-18-CE30-0001-01 and the European Union’s Horizon 2020 research and innovation programme under grant agreement No 829044.
C.R. is supported by the Marie Sklodowska-Curie program under EC Grant agreement No. 751859. O.P. is supported by an FPU predoctoral contract from MINECO No. FPU16/05460 and the Spanish MECD Grant No. FIS2014-57432-P.

\bibliography{HOTI_CD.bib}

\begin{thebibliography}{75}%
\makeatletter
\providecommand \@ifxundefined [1]{%
 \@ifx{#1\undefined}
}%
\providecommand \@ifnum [1]{%
 \ifnum #1\expandafter \@firstoftwo
 \else \expandafter \@secondoftwo
 \fi
}%
\providecommand \@ifx [1]{%
 \ifx #1\expandafter \@firstoftwo
 \else \expandafter \@secondoftwo
 \fi
}%
\providecommand \natexlab [1]{#1}%
\providecommand \enquote  [1]{``#1''}%
\providecommand \bibnamefont  [1]{#1}%
\providecommand \bibfnamefont [1]{#1}%
\providecommand \citenamefont [1]{#1}%
\providecommand \href@noop [0]{\@secondoftwo}%
\providecommand \href [0]{\begingroup \@sanitize@url \@href}%
\providecommand \@href[1]{\@@startlink{#1}\@@href}%
\providecommand \@@href[1]{\endgroup#1\@@endlink}%
\providecommand \@sanitize@url [0]{\catcode `\\12\catcode `\$12\catcode
  `\&12\catcode `\#12\catcode `\^12\catcode `\_12\catcode `\%12\relax}%
\providecommand \@@startlink[1]{}%
\providecommand \@@endlink[0]{}%
\providecommand \url  [0]{\begingroup\@sanitize@url \@url }%
\providecommand \@url [1]{\endgroup\@href {#1}{\urlprefix }}%
\providecommand \urlprefix  [0]{URL }%
\providecommand \Eprint [0]{\href }%
\providecommand \doibase [0]{http://dx.doi.org/}%
\providecommand \selectlanguage [0]{\@gobble}%
\providecommand \bibinfo  [0]{\@secondoftwo}%
\providecommand \bibfield  [0]{\@secondoftwo}%
\providecommand \translation [1]{[#1]}%
\providecommand \BibitemOpen [0]{}%
\providecommand \bibitemStop [0]{}%
\providecommand \bibitemNoStop [0]{.\EOS\space}%
\providecommand \EOS [0]{\spacefactor3000\relax}%
\providecommand \BibitemShut  [1]{\csname bibitem#1\endcsname}%
\let\auto@bib@innerbib\@empty
%</preamble>
\bibitem [{\citenamefont {Wu}\ \emph {et~al.}(2016)\citenamefont {Wu},
  \citenamefont {Salehi}, \citenamefont {Koirala}, \citenamefont {Moon},
  \citenamefont {Oh},\ and\ \citenamefont {Armitage}}]{wu2016quantized}%
  \BibitemOpen
  \bibfield  {author} {\bibinfo {author} {\bibfnamefont {L.}~\bibnamefont
  {Wu}}, \bibinfo {author} {\bibfnamefont {M.}~\bibnamefont {Salehi}}, \bibinfo
  {author} {\bibfnamefont {N.}~\bibnamefont {Koirala}}, \bibinfo {author}
  {\bibfnamefont {J.}~\bibnamefont {Moon}}, \bibinfo {author} {\bibfnamefont
  {S.}~\bibnamefont {Oh}}, \ and\ \bibinfo {author} {\bibfnamefont
  {N.}~\bibnamefont {Armitage}},\ }\href@noop {} {\bibfield  {journal}
  {\bibinfo  {journal} {Science}\ }\textbf {\bibinfo {volume} {354}},\ \bibinfo
  {pages} {1124} (\bibinfo {year} {2016})}\BibitemShut {NoStop}%
\bibitem [{\citenamefont {Dziom}\ \emph {et~al.}(2017)\citenamefont {Dziom},
  \citenamefont {Shuvaev}, \citenamefont {Pimenov}, \citenamefont {Astakhov},
  \citenamefont {Ames}, \citenamefont {Bendias}, \citenamefont {B{\"o}ttcher},
  \citenamefont {Tkachov}, \citenamefont {Hankiewicz}, \citenamefont
  {Br{\"u}ne}, \citenamefont {Buhmann},\ and\ \citenamefont
  {Molenkamp}}]{Dziom:2017uh}%
  \BibitemOpen
  \bibfield  {author} {\bibinfo {author} {\bibfnamefont {V.}~\bibnamefont
  {Dziom}}, \bibinfo {author} {\bibfnamefont {A.}~\bibnamefont {Shuvaev}},
  \bibinfo {author} {\bibfnamefont {A.}~\bibnamefont {Pimenov}}, \bibinfo
  {author} {\bibfnamefont {G.~V.}\ \bibnamefont {Astakhov}}, \bibinfo {author}
  {\bibfnamefont {C.}~\bibnamefont {Ames}}, \bibinfo {author} {\bibfnamefont
  {K.}~\bibnamefont {Bendias}}, \bibinfo {author} {\bibfnamefont
  {J.}~\bibnamefont {B{\"o}ttcher}}, \bibinfo {author} {\bibfnamefont
  {G.}~\bibnamefont {Tkachov}}, \bibinfo {author} {\bibfnamefont {E.~M.}\
  \bibnamefont {Hankiewicz}}, \bibinfo {author} {\bibfnamefont
  {C.}~\bibnamefont {Br{\"u}ne}}, \bibinfo {author} {\bibfnamefont
  {H.}~\bibnamefont {Buhmann}}, \ and\ \bibinfo {author} {\bibfnamefont
  {L.~W.}\ \bibnamefont {Molenkamp}},\ }\href
  {https://doi.org/10.1038/ncomms15197} {\bibfield  {journal} {\bibinfo
  {journal} {Nature Communications}\ }\textbf {\bibinfo {volume} {8}},\
  \bibinfo {pages} {15197} (\bibinfo {year} {2017})}\BibitemShut {NoStop}%
\bibitem [{\citenamefont {Rees}\ \emph {et~al.}(2019)\citenamefont {Rees},
  \citenamefont {Manna}, \citenamefont {Lu}, \citenamefont {Morimoto},
  \citenamefont {Borrmann}, \citenamefont {Felser}, \citenamefont {Moore},
  \citenamefont {Torchinsky},\ and\ \citenamefont {Orenstein}}]{Rees:2019ue}%
  \BibitemOpen
  \bibfield  {author} {\bibinfo {author} {\bibfnamefont {D.}~\bibnamefont
  {Rees}}, \bibinfo {author} {\bibfnamefont {K.}~\bibnamefont {Manna}},
  \bibinfo {author} {\bibfnamefont {B.}~\bibnamefont {Lu}}, \bibinfo {author}
  {\bibfnamefont {T.}~\bibnamefont {Morimoto}}, \bibinfo {author}
  {\bibfnamefont {H.}~\bibnamefont {Borrmann}}, \bibinfo {author}
  {\bibfnamefont {C.}~\bibnamefont {Felser}}, \bibinfo {author} {\bibfnamefont
  {J.~E.}\ \bibnamefont {Moore}}, \bibinfo {author} {\bibfnamefont {D.~H.}\
  \bibnamefont {Torchinsky}}, \ and\ \bibinfo {author} {\bibfnamefont
  {J.}~\bibnamefont {Orenstein}},\ }\href@noop {} {\bibfield  {journal}
  {\bibinfo  {journal} {arXiv.org}\ } (\bibinfo {year} {2019})},\ \Eprint
  {http://arxiv.org/abs/1902.03230v1} {1902.03230v1} \BibitemShut {NoStop}%
\bibitem [{\citenamefont {Benalcazar}\ \emph
  {et~al.}(2017{\natexlab{a}})\citenamefont {Benalcazar}, \citenamefont
  {Bernevig},\ and\ \citenamefont {Hughes}}]{Benalcazar}%
  \BibitemOpen
  \bibfield  {author} {\bibinfo {author} {\bibfnamefont {W.~A.}\ \bibnamefont
  {Benalcazar}}, \bibinfo {author} {\bibfnamefont {B.~A.}\ \bibnamefont
  {Bernevig}}, \ and\ \bibinfo {author} {\bibfnamefont {T.~L.}\ \bibnamefont
  {Hughes}},\ }\href@noop {} {\bibfield  {journal} {\bibinfo  {journal}
  {Science}\ }\textbf {\bibinfo {volume} {357}},\ \bibinfo {pages} {61}
  (\bibinfo {year} {2017}{\natexlab{a}})}\BibitemShut {NoStop}%
\bibitem [{\citenamefont {Benalcazar}\ \emph
  {et~al.}(2017{\natexlab{b}})\citenamefont {Benalcazar}, \citenamefont
  {Bernevig},\ and\ \citenamefont {Hughes}}]{Benalcazar:2017cn}%
  \BibitemOpen
  \bibfield  {author} {\bibinfo {author} {\bibfnamefont {W.~A.}\ \bibnamefont
  {Benalcazar}}, \bibinfo {author} {\bibfnamefont {B.~A.}\ \bibnamefont
  {Bernevig}}, \ and\ \bibinfo {author} {\bibfnamefont {T.~L.}\ \bibnamefont
  {Hughes}},\ }\href@noop {} {\bibfield  {journal} {\bibinfo  {journal}
  {Physical Review B}\ }\textbf {\bibinfo {volume} {96}},\ \bibinfo {pages}
  {245115} (\bibinfo {year} {2017}{\natexlab{b}})}\BibitemShut {NoStop}%
\bibitem [{\citenamefont {Schindler}\ \emph
  {et~al.}(2018{\natexlab{a}})\citenamefont {Schindler}, \citenamefont {Cook},
  \citenamefont {Vergniory}, \citenamefont {Wang}, \citenamefont {Parkin},
  \citenamefont {Bernevig},\ and\ \citenamefont
  {Neupert}}]{schindler2018higher}%
  \BibitemOpen
  \bibfield  {author} {\bibinfo {author} {\bibfnamefont {F.}~\bibnamefont
  {Schindler}}, \bibinfo {author} {\bibfnamefont {A.~M.}\ \bibnamefont {Cook}},
  \bibinfo {author} {\bibfnamefont {M.~G.}\ \bibnamefont {Vergniory}}, \bibinfo
  {author} {\bibfnamefont {Z.}~\bibnamefont {Wang}}, \bibinfo {author}
  {\bibfnamefont {S.~S.}\ \bibnamefont {Parkin}}, \bibinfo {author}
  {\bibfnamefont {B.~A.}\ \bibnamefont {Bernevig}}, \ and\ \bibinfo {author}
  {\bibfnamefont {T.}~\bibnamefont {Neupert}},\ }\href@noop {} {\bibfield
  {journal} {\bibinfo  {journal} {Science advances}\ }\textbf {\bibinfo
  {volume} {4}},\ \bibinfo {pages} {eaat0346} (\bibinfo {year}
  {2018}{\natexlab{a}})}\BibitemShut {NoStop}%
\bibitem [{\citenamefont {Song}\ \emph {et~al.}(2017)\citenamefont {Song},
  \citenamefont {Fang},\ and\ \citenamefont {Fang}}]{Song:2017eva}%
  \BibitemOpen
  \bibfield  {author} {\bibinfo {author} {\bibfnamefont {Z.}~\bibnamefont
  {Song}}, \bibinfo {author} {\bibfnamefont {Z.}~\bibnamefont {Fang}}, \ and\
  \bibinfo {author} {\bibfnamefont {C.}~\bibnamefont {Fang}},\ }\href@noop {}
  {\bibfield  {journal} {\bibinfo  {journal} {Physical Review Letters}\
  }\textbf {\bibinfo {volume} {119}},\ \bibinfo {pages} {246402} (\bibinfo
  {year} {2017})}\BibitemShut {NoStop}%
\bibitem [{\citenamefont {Sitte}\ \emph {et~al.}(2012)\citenamefont {Sitte},
  \citenamefont {Rosch}, \citenamefont {Altman},\ and\ \citenamefont
  {Fritz}}]{Sitte:2012ib}%
  \BibitemOpen
  \bibfield  {author} {\bibinfo {author} {\bibfnamefont {M.}~\bibnamefont
  {Sitte}}, \bibinfo {author} {\bibfnamefont {A.}~\bibnamefont {Rosch}},
  \bibinfo {author} {\bibfnamefont {E.}~\bibnamefont {Altman}}, \ and\ \bibinfo
  {author} {\bibfnamefont {L.}~\bibnamefont {Fritz}},\ }\href@noop {}
  {\bibfield  {journal} {\bibinfo  {journal} {Physical Review Letters}\
  }\textbf {\bibinfo {volume} {108}},\ \bibinfo {pages} {126807} (\bibinfo
  {year} {2012})}\BibitemShut {NoStop}%
\bibitem [{\citenamefont {Zhang}\ \emph {et~al.}(2013)\citenamefont {Zhang},
  \citenamefont {Kane},\ and\ \citenamefont {Mele}}]{Zhang2013}%
  \BibitemOpen
  \bibfield  {author} {\bibinfo {author} {\bibfnamefont {F.}~\bibnamefont
  {Zhang}}, \bibinfo {author} {\bibfnamefont {C.~L.}\ \bibnamefont {Kane}}, \
  and\ \bibinfo {author} {\bibfnamefont {E.~J.}\ \bibnamefont {Mele}},\ }\href
  {\doibase 10.1103/PhysRevLett.110.046404} {\bibfield  {journal} {\bibinfo
  {journal} {Phys. Rev. Lett.}\ }\textbf {\bibinfo {volume} {110}},\ \bibinfo
  {pages} {046404} (\bibinfo {year} {2013})}\BibitemShut {NoStop}%
\bibitem [{\citenamefont {Langbehn}\ \emph {et~al.}(2017)\citenamefont
  {Langbehn}, \citenamefont {Peng}, \citenamefont {Trifunovic}, \citenamefont
  {von Oppen},\ and\ \citenamefont {Brouwer}}]{PhysRevLett.119.246401}%
  \BibitemOpen
  \bibfield  {author} {\bibinfo {author} {\bibfnamefont {J.}~\bibnamefont
  {Langbehn}}, \bibinfo {author} {\bibfnamefont {Y.}~\bibnamefont {Peng}},
  \bibinfo {author} {\bibfnamefont {L.}~\bibnamefont {Trifunovic}}, \bibinfo
  {author} {\bibfnamefont {F.}~\bibnamefont {von Oppen}}, \ and\ \bibinfo
  {author} {\bibfnamefont {P.~W.}\ \bibnamefont {Brouwer}},\ }\href {\doibase
  10.1103/PhysRevLett.119.246401} {\bibfield  {journal} {\bibinfo  {journal}
  {Phys. Rev. Lett.}\ }\textbf {\bibinfo {volume} {119}},\ \bibinfo {pages}
  {246401} (\bibinfo {year} {2017})}\BibitemShut {NoStop}%
\bibitem [{\citenamefont {Liu}\ and\ \citenamefont
  {Wakabayashi}(2017)}]{Wakabayashi2017}%
  \BibitemOpen
  \bibfield  {author} {\bibinfo {author} {\bibfnamefont {F.}~\bibnamefont
  {Liu}}\ and\ \bibinfo {author} {\bibfnamefont {K.}~\bibnamefont
  {Wakabayashi}},\ }\href {\doibase 10.1103/PhysRevLett.118.076803} {\bibfield
  {journal} {\bibinfo  {journal} {Phys. Rev. Lett.}\ }\textbf {\bibinfo
  {volume} {118}},\ \bibinfo {pages} {076803} (\bibinfo {year}
  {2017})}\BibitemShut {NoStop}%
\bibitem [{\citenamefont {Ezawa}(2018{\natexlab{a}})}]{Ezawa:2018gt}%
  \BibitemOpen
  \bibfield  {author} {\bibinfo {author} {\bibfnamefont {M.}~\bibnamefont
  {Ezawa}},\ }\href@noop {} {\bibfield  {journal} {\bibinfo  {journal}
  {Physical Review B}\ }\textbf {\bibinfo {volume} {98}},\ \bibinfo {pages}
  {045125} (\bibinfo {year} {2018}{\natexlab{a}})}\BibitemShut {NoStop}%
\bibitem [{\citenamefont {Ezawa}(2018{\natexlab{b}})}]{ezawa2018magnetic}%
  \BibitemOpen
  \bibfield  {author} {\bibinfo {author} {\bibfnamefont {M.}~\bibnamefont
  {Ezawa}},\ }\href@noop {} {\bibfield  {journal} {\bibinfo  {journal}
  {Physical Review B}\ }\textbf {\bibinfo {volume} {97}},\ \bibinfo {pages}
  {155305} (\bibinfo {year} {2018}{\natexlab{b}})}\BibitemShut {NoStop}%
\bibitem [{\citenamefont {Ahn}\ and\ \citenamefont {Yang}(2019)}]{Ahn:uc}%
  \BibitemOpen
  \bibfield  {author} {\bibinfo {author} {\bibfnamefont {J.}~\bibnamefont
  {Ahn}}\ and\ \bibinfo {author} {\bibfnamefont {B.-J.}\ \bibnamefont {Yang}},\
  }\href {\doibase 10.1103/PhysRevB.99.235125} {\bibfield  {journal} {\bibinfo
  {journal} {Phys. Rev. B}\ }\textbf {\bibinfo {volume} {99}},\ \bibinfo
  {pages} {235125} (\bibinfo {year} {2019})}\BibitemShut {NoStop}%
\bibitem [{\citenamefont {{Wang}}\ \emph {et~al.}(2018)\citenamefont {{Wang}},
  \citenamefont {{Wieder}}, \citenamefont {{Li}}, \citenamefont {{Yan}},\ and\
  \citenamefont {{Bernevig}}}]{Wang2018}%
  \BibitemOpen
  \bibfield  {author} {\bibinfo {author} {\bibfnamefont {Z.}~\bibnamefont
  {{Wang}}}, \bibinfo {author} {\bibfnamefont {B.~J.}\ \bibnamefont
  {{Wieder}}}, \bibinfo {author} {\bibfnamefont {J.}~\bibnamefont {{Li}}},
  \bibinfo {author} {\bibfnamefont {B.}~\bibnamefont {{Yan}}}, \ and\ \bibinfo
  {author} {\bibfnamefont {B.~A.}\ \bibnamefont {{Bernevig}}},\ }\href@noop {}
  {\bibfield  {journal} {\bibinfo  {journal} {arXiv}\ ,\ \bibinfo {eid}
  {arXiv:1806.11116}} (\bibinfo {year} {2018})}\BibitemShut {NoStop}%
\bibitem [{\citenamefont {Wieder}\ and\ \citenamefont
  {Bernevig}(2018)}]{Wieder:2018qpe}%
  \BibitemOpen
  \bibfield  {author} {\bibinfo {author} {\bibfnamefont {B.~J.}\ \bibnamefont
  {Wieder}}\ and\ \bibinfo {author} {\bibfnamefont {B.~A.}\ \bibnamefont
  {Bernevig}},\ }\href@noop {} {\bibfield  {journal} {\bibinfo  {journal}
  {arXiv}\ } (\bibinfo {year} {2018})},\ \Eprint
  {http://arxiv.org/abs/1810.02373v1} {1810.02373v1} \BibitemShut {NoStop}%
\bibitem [{\citenamefont {Lin}\ and\ \citenamefont {Hughes}(2018)}]{Lin2017}%
  \BibitemOpen
  \bibfield  {author} {\bibinfo {author} {\bibfnamefont {M.}~\bibnamefont
  {Lin}}\ and\ \bibinfo {author} {\bibfnamefont {T.~L.}\ \bibnamefont
  {Hughes}},\ }\href {\doibase 10.1103/PhysRevB.98.241103} {\bibfield
  {journal} {\bibinfo  {journal} {Phys. Rev. B}\ }\textbf {\bibinfo {volume}
  {98}},\ \bibinfo {pages} {241103} (\bibinfo {year} {2018})}\BibitemShut
  {NoStop}%
\bibitem [{\citenamefont {Khalaf}(2018)}]{khalaf2018higher}%
  \BibitemOpen
  \bibfield  {author} {\bibinfo {author} {\bibfnamefont {E.}~\bibnamefont
  {Khalaf}},\ }\href@noop {} {\bibfield  {journal} {\bibinfo  {journal}
  {Physical Review B}\ }\textbf {\bibinfo {volume} {97}},\ \bibinfo {pages}
  {205136} (\bibinfo {year} {2018})}\BibitemShut {NoStop}%
\bibitem [{\citenamefont {Varnava}\ and\ \citenamefont
  {Vanderbilt}(2018)}]{varnava2018surfaces}%
  \BibitemOpen
  \bibfield  {author} {\bibinfo {author} {\bibfnamefont {N.}~\bibnamefont
  {Varnava}}\ and\ \bibinfo {author} {\bibfnamefont {D.}~\bibnamefont
  {Vanderbilt}},\ }\href@noop {} {\bibfield  {journal} {\bibinfo  {journal}
  {Physical Review B}\ }\textbf {\bibinfo {volume} {98}},\ \bibinfo {pages}
  {245117} (\bibinfo {year} {2018})}\BibitemShut {NoStop}%
\bibitem [{\citenamefont {Kooi}\ \emph {et~al.}(2018)\citenamefont {Kooi},
  \citenamefont {van Miert},\ and\ \citenamefont {Ortix}}]{Kooi:2018cr}%
  \BibitemOpen
  \bibfield  {author} {\bibinfo {author} {\bibfnamefont {S.~H.}\ \bibnamefont
  {Kooi}}, \bibinfo {author} {\bibfnamefont {G.}~\bibnamefont {van Miert}}, \
  and\ \bibinfo {author} {\bibfnamefont {C.}~\bibnamefont {Ortix}},\
  }\href@noop {} {\bibfield  {journal} {\bibinfo  {journal} {Physical Review
  B}\ }\textbf {\bibinfo {volume} {98}},\ \bibinfo {pages} {245102} (\bibinfo
  {year} {2018})}\BibitemShut {NoStop}%
\bibitem [{\citenamefont {van Miert}\ and\ \citenamefont
  {Ortix}(2018)}]{vanMiert2018}%
  \BibitemOpen
  \bibfield  {author} {\bibinfo {author} {\bibfnamefont {G.}~\bibnamefont {van
  Miert}}\ and\ \bibinfo {author} {\bibfnamefont {C.}~\bibnamefont {Ortix}},\
  }\href {\doibase 10.1103/PhysRevB.98.081110} {\bibfield  {journal} {\bibinfo
  {journal} {Phys. Rev. B}\ }\textbf {\bibinfo {volume} {98}},\ \bibinfo
  {pages} {081110} (\bibinfo {year} {2018})}\BibitemShut {NoStop}%
\bibitem [{\citenamefont {Franca}\ \emph {et~al.}(2018)\citenamefont {Franca},
  \citenamefont {van~den Brink},\ and\ \citenamefont {Fulga}}]{Franca2018}%
  \BibitemOpen
  \bibfield  {author} {\bibinfo {author} {\bibfnamefont {S.}~\bibnamefont
  {Franca}}, \bibinfo {author} {\bibfnamefont {J.}~\bibnamefont {van~den
  Brink}}, \ and\ \bibinfo {author} {\bibfnamefont {I.~C.}\ \bibnamefont
  {Fulga}},\ }\href {\doibase 10.1103/PhysRevB.98.201114} {\bibfield  {journal}
  {\bibinfo  {journal} {Phys. Rev. B}\ }\textbf {\bibinfo {volume} {98}},\
  \bibinfo {pages} {201114} (\bibinfo {year} {2018})}\BibitemShut {NoStop}%
\bibitem [{\citenamefont {Trifunovic}\ and\ \citenamefont
  {Brouwer}(2019)}]{Trifunovic2019}%
  \BibitemOpen
  \bibfield  {author} {\bibinfo {author} {\bibfnamefont {L.}~\bibnamefont
  {Trifunovic}}\ and\ \bibinfo {author} {\bibfnamefont {P.~W.}\ \bibnamefont
  {Brouwer}},\ }\href {\doibase 10.1103/PhysRevX.9.011012} {\bibfield
  {journal} {\bibinfo  {journal} {Phys. Rev. X}\ }\textbf {\bibinfo {volume}
  {9}},\ \bibinfo {pages} {011012} (\bibinfo {year} {2019})}\BibitemShut
  {NoStop}%
\bibitem [{\citenamefont {Matsugatani}\ and\ \citenamefont
  {Watanabe}(2018)}]{Matsugatani:2018jb}%
  \BibitemOpen
  \bibfield  {author} {\bibinfo {author} {\bibfnamefont {A.}~\bibnamefont
  {Matsugatani}}\ and\ \bibinfo {author} {\bibfnamefont {H.}~\bibnamefont
  {Watanabe}},\ }\href@noop {} {\bibfield  {journal} {\bibinfo  {journal}
  {Physical Review B}\ }\textbf {\bibinfo {volume} {98}},\ \bibinfo {pages}
  {205129} (\bibinfo {year} {2018})}\BibitemShut {NoStop}%
\bibitem [{\citenamefont {Song}\ \emph {et~al.}(2018)\citenamefont {Song},
  \citenamefont {Zhang}, \citenamefont {Fang},\ and\ \citenamefont
  {Fang}}]{Song:2018cj}%
  \BibitemOpen
  \bibfield  {author} {\bibinfo {author} {\bibfnamefont {Z.}~\bibnamefont
  {Song}}, \bibinfo {author} {\bibfnamefont {T.}~\bibnamefont {Zhang}},
  \bibinfo {author} {\bibfnamefont {Z.}~\bibnamefont {Fang}}, \ and\ \bibinfo
  {author} {\bibfnamefont {C.}~\bibnamefont {Fang}},\ }\href@noop {} {\bibfield
   {journal} {\bibinfo  {journal} {Nature Communications}\ }\textbf {\bibinfo
  {volume} {9}},\ \bibinfo {pages} {3530} (\bibinfo {year} {2018})}\BibitemShut
  {NoStop}%
\bibitem [{\citenamefont {Schindler}\ \emph
  {et~al.}(2018{\natexlab{b}})\citenamefont {Schindler}, \citenamefont {Wang},
  \citenamefont {Vergniory}, \citenamefont {Cook}, \citenamefont {Murani},
  \citenamefont {Sengupta}, \citenamefont {Kasumov}, \citenamefont {Deblock},
  \citenamefont {Jeon}, \citenamefont {Drozdov}, \citenamefont {Bouchiat},
  \citenamefont {Gu{\'e}ron}, \citenamefont {Yazdani}, \citenamefont
  {Bernevig},\ and\ \citenamefont {Neupert}}]{Schindler2018bism}%
  \BibitemOpen
  \bibfield  {author} {\bibinfo {author} {\bibfnamefont {F.}~\bibnamefont
  {Schindler}}, \bibinfo {author} {\bibfnamefont {Z.}~\bibnamefont {Wang}},
  \bibinfo {author} {\bibfnamefont {M.~G.}\ \bibnamefont {Vergniory}}, \bibinfo
  {author} {\bibfnamefont {A.~M.}\ \bibnamefont {Cook}}, \bibinfo {author}
  {\bibfnamefont {A.}~\bibnamefont {Murani}}, \bibinfo {author} {\bibfnamefont
  {S.}~\bibnamefont {Sengupta}}, \bibinfo {author} {\bibfnamefont {A.~Y.}\
  \bibnamefont {Kasumov}}, \bibinfo {author} {\bibfnamefont {R.}~\bibnamefont
  {Deblock}}, \bibinfo {author} {\bibfnamefont {S.}~\bibnamefont {Jeon}},
  \bibinfo {author} {\bibfnamefont {I.}~\bibnamefont {Drozdov}}, \bibinfo
  {author} {\bibfnamefont {H.}~\bibnamefont {Bouchiat}}, \bibinfo {author}
  {\bibfnamefont {S.}~\bibnamefont {Gu{\'e}ron}}, \bibinfo {author}
  {\bibfnamefont {A.}~\bibnamefont {Yazdani}}, \bibinfo {author} {\bibfnamefont
  {B.~A.}\ \bibnamefont {Bernevig}}, \ and\ \bibinfo {author} {\bibfnamefont
  {T.}~\bibnamefont {Neupert}},\ }\href {\doibase 10.1038/s41567-018-0224-7}
  {\bibfield  {journal} {\bibinfo  {journal} {Nature Physics}\ }\textbf
  {\bibinfo {volume} {14}},\ \bibinfo {pages} {918} (\bibinfo {year}
  {2018}{\natexlab{b}})}\BibitemShut {NoStop}%
\bibitem [{\citenamefont {Sessi}\ \emph {et~al.}(2016)\citenamefont {Sessi},
  \citenamefont {Di~Sante}, \citenamefont {Szczerbakow}, \citenamefont {Glott},
  \citenamefont {Wilfert}, \citenamefont {Schmidt}, \citenamefont {Bathon},
  \citenamefont {Dziawa}, \citenamefont {Greiter}, \citenamefont {Neupert},
  \citenamefont {Sangiovanni}, \citenamefont {Story}, \citenamefont {Thomale},\
  and\ \citenamefont {Bode}}]{SessiScience2016}%
  \BibitemOpen
  \bibfield  {author} {\bibinfo {author} {\bibfnamefont {P.}~\bibnamefont
  {Sessi}}, \bibinfo {author} {\bibfnamefont {D.}~\bibnamefont {Di~Sante}},
  \bibinfo {author} {\bibfnamefont {A.}~\bibnamefont {Szczerbakow}}, \bibinfo
  {author} {\bibfnamefont {F.}~\bibnamefont {Glott}}, \bibinfo {author}
  {\bibfnamefont {S.}~\bibnamefont {Wilfert}}, \bibinfo {author} {\bibfnamefont
  {H.}~\bibnamefont {Schmidt}}, \bibinfo {author} {\bibfnamefont
  {T.}~\bibnamefont {Bathon}}, \bibinfo {author} {\bibfnamefont
  {P.}~\bibnamefont {Dziawa}}, \bibinfo {author} {\bibfnamefont
  {M.}~\bibnamefont {Greiter}}, \bibinfo {author} {\bibfnamefont
  {T.}~\bibnamefont {Neupert}}, \bibinfo {author} {\bibfnamefont
  {G.}~\bibnamefont {Sangiovanni}}, \bibinfo {author} {\bibfnamefont
  {T.}~\bibnamefont {Story}}, \bibinfo {author} {\bibfnamefont
  {R.}~\bibnamefont {Thomale}}, \ and\ \bibinfo {author} {\bibfnamefont
  {M.}~\bibnamefont {Bode}},\ }\href {\doibase 10.1126/science.aah6233}
  {\bibfield  {journal} {\bibinfo  {journal} {Science}\ }\textbf {\bibinfo
  {volume} {354}},\ \bibinfo {pages} {1269} (\bibinfo {year} {2016})},\ \Eprint
  {http://arxiv.org/abs/https://science.sciencemag.org/content/354/6317/1269.full.pdf}
  {https://science.sciencemag.org/content/354/6317/1269.full.pdf} \BibitemShut
  {NoStop}%
\bibitem [{\citenamefont {Hasan}\ and\ \citenamefont
  {Kane}(2010)}]{HasanKane2010}%
  \BibitemOpen
  \bibfield  {author} {\bibinfo {author} {\bibfnamefont {M.~Z.}\ \bibnamefont
  {Hasan}}\ and\ \bibinfo {author} {\bibfnamefont {C.~L.}\ \bibnamefont
  {Kane}},\ }\href {\doibase 10.1103/RevModPhys.82.3045} {\bibfield  {journal}
  {\bibinfo  {journal} {Rev. Mod. Phys.}\ }\textbf {\bibinfo {volume} {82}},\
  \bibinfo {pages} {3045} (\bibinfo {year} {2010})}\BibitemShut {NoStop}%
\bibitem [{\citenamefont {Qi}\ and\ \citenamefont {Zhang}(2011)}]{Qi2011}%
  \BibitemOpen
  \bibfield  {author} {\bibinfo {author} {\bibfnamefont {X.-L.}\ \bibnamefont
  {Qi}}\ and\ \bibinfo {author} {\bibfnamefont {S.-C.}\ \bibnamefont {Zhang}},\
  }\href {\doibase 10.1103/RevModPhys.83.1057} {\bibfield  {journal} {\bibinfo
  {journal} {Rev. Mod. Phys.}\ }\textbf {\bibinfo {volume} {83}},\ \bibinfo
  {pages} {1057} (\bibinfo {year} {2011})}\BibitemShut {NoStop}%
\bibitem [{\citenamefont {Wang}\ \emph {et~al.}(2011)\citenamefont {Wang},
  \citenamefont {Hsieh}, \citenamefont {Pilon}, \citenamefont {Fu},
  \citenamefont {Gardner}, \citenamefont {Lee},\ and\ \citenamefont
  {Gedik}}]{WangHsieh2011}%
  \BibitemOpen
  \bibfield  {author} {\bibinfo {author} {\bibfnamefont {Y.~H.}\ \bibnamefont
  {Wang}}, \bibinfo {author} {\bibfnamefont {D.}~\bibnamefont {Hsieh}},
  \bibinfo {author} {\bibfnamefont {D.}~\bibnamefont {Pilon}}, \bibinfo
  {author} {\bibfnamefont {L.}~\bibnamefont {Fu}}, \bibinfo {author}
  {\bibfnamefont {D.~R.}\ \bibnamefont {Gardner}}, \bibinfo {author}
  {\bibfnamefont {Y.~S.}\ \bibnamefont {Lee}}, \ and\ \bibinfo {author}
  {\bibfnamefont {N.}~\bibnamefont {Gedik}},\ }\href {\doibase
  10.1103/PhysRevLett.107.207602} {\bibfield  {journal} {\bibinfo  {journal}
  {Phys. Rev. Lett.}\ }\textbf {\bibinfo {volume} {107}},\ \bibinfo {pages}
  {207602} (\bibinfo {year} {2011})}\BibitemShut {NoStop}%
\bibitem [{\citenamefont {Goldman}\ \emph {et~al.}(2012)\citenamefont
  {Goldman}, \citenamefont {Beugnon},\ and\ \citenamefont
  {Gerbier}}]{Goldman2012}%
  \BibitemOpen
  \bibfield  {author} {\bibinfo {author} {\bibfnamefont {N.}~\bibnamefont
  {Goldman}}, \bibinfo {author} {\bibfnamefont {J.}~\bibnamefont {Beugnon}}, \
  and\ \bibinfo {author} {\bibfnamefont {F.}~\bibnamefont {Gerbier}},\ }\href
  {\doibase 10.1103/PhysRevLett.108.255303} {\bibfield  {journal} {\bibinfo
  {journal} {Phys. Rev. Lett.}\ }\textbf {\bibinfo {volume} {108}},\ \bibinfo
  {pages} {255303} (\bibinfo {year} {2012})}\BibitemShut {NoStop}%
\bibitem [{\citenamefont {Souza}\ and\ \citenamefont
  {Vanderbilt}(2008)}]{Souza2008}%
  \BibitemOpen
  \bibfield  {author} {\bibinfo {author} {\bibfnamefont {I.}~\bibnamefont
  {Souza}}\ and\ \bibinfo {author} {\bibfnamefont {D.}~\bibnamefont
  {Vanderbilt}},\ }\href@noop {} {\bibfield  {journal} {\bibinfo  {journal}
  {Physical Review B}\ }\textbf {\bibinfo {volume} {77}},\ \bibinfo {pages}
  {054438} (\bibinfo {year} {2008})}\BibitemShut {NoStop}%
\bibitem [{\citenamefont {Wang}\ and\ \citenamefont
  {Gedik}(2013)}]{WangGedik2013}%
  \BibitemOpen
  \bibfield  {author} {\bibinfo {author} {\bibfnamefont {Y.}~\bibnamefont
  {Wang}}\ and\ \bibinfo {author} {\bibfnamefont {N.}~\bibnamefont {Gedik}},\
  }\href {\doibase 10.1002/pssr.201206458} {\bibfield  {journal} {\bibinfo
  {journal} {physica status solidi (RRL) – Rapid Research Letters}\ }\textbf
  {\bibinfo {volume} {7}},\ \bibinfo {pages} {64} (\bibinfo {year}
  {2013})}\BibitemShut {NoStop}%
\bibitem [{\citenamefont {Sch\"uler}\ and\ \citenamefont
  {Werner}(2017)}]{Schuler2017}%
  \BibitemOpen
  \bibfield  {author} {\bibinfo {author} {\bibfnamefont {M.}~\bibnamefont
  {Sch\"uler}}\ and\ \bibinfo {author} {\bibfnamefont {P.}~\bibnamefont
  {Werner}},\ }\href {\doibase 10.1103/PhysRevB.96.155122} {\bibfield
  {journal} {\bibinfo  {journal} {Phys. Rev. B}\ }\textbf {\bibinfo {volume}
  {96}},\ \bibinfo {pages} {155122} (\bibinfo {year} {2017})}\BibitemShut
  {NoStop}%
\bibitem [{\citenamefont {de~Juan}\ \emph {et~al.}(2017)\citenamefont
  {de~Juan}, \citenamefont {Grushin}, \citenamefont {Morimoto},\ and\
  \citenamefont {Moore}}]{de_juan_quantized_2017}%
  \BibitemOpen
  \bibfield  {author} {\bibinfo {author} {\bibfnamefont {F.}~\bibnamefont
  {de~Juan}}, \bibinfo {author} {\bibfnamefont {A.~G.}\ \bibnamefont
  {Grushin}}, \bibinfo {author} {\bibfnamefont {T.}~\bibnamefont {Morimoto}}, \
  and\ \bibinfo {author} {\bibfnamefont {J.~E.}\ \bibnamefont {Moore}},\ }\href
  {\doibase 10.1038/ncomms15995} {\bibfield  {journal} {\bibinfo  {journal}
  {Nature Communications}\ }\textbf {\bibinfo {volume} {8}},\ \bibinfo {pages}
  {15995} (\bibinfo {year} {2017})},\ \bibinfo {note} {arXiv:
  1611.05887}\BibitemShut {NoStop}%
\bibitem [{\citenamefont {Liu}\ \emph {et~al.}(2018)\citenamefont {Liu},
  \citenamefont {Yang},\ and\ \citenamefont {Zhang}}]{LiuYing2018}%
  \BibitemOpen
  \bibfield  {author} {\bibinfo {author} {\bibfnamefont {Y.}~\bibnamefont
  {Liu}}, \bibinfo {author} {\bibfnamefont {S.~A.}\ \bibnamefont {Yang}}, \
  and\ \bibinfo {author} {\bibfnamefont {F.}~\bibnamefont {Zhang}},\ }\href
  {\doibase 10.1103/PhysRevB.97.035153} {\bibfield  {journal} {\bibinfo
  {journal} {Phys. Rev. B}\ }\textbf {\bibinfo {volume} {97}},\ \bibinfo
  {pages} {035153} (\bibinfo {year} {2018})}\BibitemShut {NoStop}%
\bibitem [{\citenamefont {Tran}\ \emph {et~al.}(2017)\citenamefont {Tran},
  \citenamefont {Dauphin}, \citenamefont {Grushin}, \citenamefont {Zoller},\
  and\ \citenamefont {Goldman}}]{Tran2017}%
  \BibitemOpen
  \bibfield  {author} {\bibinfo {author} {\bibfnamefont {D.~T.}\ \bibnamefont
  {Tran}}, \bibinfo {author} {\bibfnamefont {A.}~\bibnamefont {Dauphin}},
  \bibinfo {author} {\bibfnamefont {A.~G.}\ \bibnamefont {Grushin}}, \bibinfo
  {author} {\bibfnamefont {P.}~\bibnamefont {Zoller}}, \ and\ \bibinfo {author}
  {\bibfnamefont {N.}~\bibnamefont {Goldman}},\ }\href {\doibase
  10.1126/sciadv.1701207} {\bibfield  {journal} {\bibinfo  {journal} {Science
  Advances}\ }\textbf {\bibinfo {volume} {3}},\ \bibinfo {pages} {e1701207}
  (\bibinfo {year} {2017})}\BibitemShut {NoStop}%
\bibitem [{\citenamefont {Tran}\ \emph {et~al.}(2018)\citenamefont {Tran},
  \citenamefont {Cooper},\ and\ \citenamefont {Goldman}}]{TranCooper2018}%
  \BibitemOpen
  \bibfield  {author} {\bibinfo {author} {\bibfnamefont {D.~T.}\ \bibnamefont
  {Tran}}, \bibinfo {author} {\bibfnamefont {N.~R.}\ \bibnamefont {Cooper}}, \
  and\ \bibinfo {author} {\bibfnamefont {N.}~\bibnamefont {Goldman}},\ }\href
  {\doibase 10.1103/PhysRevA.97.061602} {\bibfield  {journal} {\bibinfo
  {journal} {Phys. Rev. A}\ }\textbf {\bibinfo {volume} {97}},\ \bibinfo
  {pages} {061602} (\bibinfo {year} {2018})}\BibitemShut {NoStop}%
\bibitem [{\citenamefont {Repellin}\ and\ \citenamefont
  {Goldman}(2019)}]{Repellin2019}%
  \BibitemOpen
  \bibfield  {author} {\bibinfo {author} {\bibfnamefont {C.}~\bibnamefont
  {Repellin}}\ and\ \bibinfo {author} {\bibfnamefont {N.}~\bibnamefont
  {Goldman}},\ }\href {\doibase 10.1103/PhysRevLett.122.166801} {\bibfield
  {journal} {\bibinfo  {journal} {Phys. Rev. Lett.}\ }\textbf {\bibinfo
  {volume} {122}},\ \bibinfo {pages} {166801} (\bibinfo {year}
  {2019})}\BibitemShut {NoStop}%
\bibitem [{\citenamefont {{Sch{\"u}ler}}\ \emph {et~al.}(2019)\citenamefont
  {{Sch{\"u}ler}}, \citenamefont {{De Giovannini}}, \citenamefont
  {{H{\"u}bener}}, \citenamefont {{Rubio}}, \citenamefont {{Sentef}},\ and\
  \citenamefont {{Werner}}}]{Schuler2019}%
  \BibitemOpen
  \bibfield  {author} {\bibinfo {author} {\bibfnamefont {M.}~\bibnamefont
  {{Sch{\"u}ler}}}, \bibinfo {author} {\bibfnamefont {U.}~\bibnamefont {{De
  Giovannini}}}, \bibinfo {author} {\bibfnamefont {H.}~\bibnamefont
  {{H{\"u}bener}}}, \bibinfo {author} {\bibfnamefont {A.}~\bibnamefont
  {{Rubio}}}, \bibinfo {author} {\bibfnamefont {M.~A.}\ \bibnamefont
  {{Sentef}}}, \ and\ \bibinfo {author} {\bibfnamefont {P.}~\bibnamefont
  {{Werner}}},\ }\href@noop {} {\bibfield  {journal} {\bibinfo  {journal}
  {arXiv}\ ,\ \bibinfo {eid} {arXiv:1905.09404}} (\bibinfo {year}
  {2019})}\BibitemShut {NoStop}%
\bibitem [{\citenamefont {Bennett}\ and\ \citenamefont
  {Stern}(1965)}]{Bennett:1965jy}%
  \BibitemOpen
  \bibfield  {author} {\bibinfo {author} {\bibfnamefont {H.~S.}\ \bibnamefont
  {Bennett}}\ and\ \bibinfo {author} {\bibfnamefont {E.~A.}\ \bibnamefont
  {Stern}},\ }\href@noop {} {\bibfield  {journal} {\bibinfo  {journal}
  {Physical Review}\ }\textbf {\bibinfo {volume} {137}},\ \bibinfo {pages}
  {A448} (\bibinfo {year} {1965})}\BibitemShut {NoStop}%
\bibitem [{\citenamefont {Asteria}\ \emph {et~al.}(2019)\citenamefont
  {Asteria}, \citenamefont {Tran}, \citenamefont {Ozawa}, \citenamefont
  {Tarnowski}, \citenamefont {Rem}, \citenamefont {Fl{\"a}schner},
  \citenamefont {Sengstock}, \citenamefont {Goldman},\ and\ \citenamefont
  {Weitenberg}}]{Asteria:2019if}%
  \BibitemOpen
  \bibfield  {author} {\bibinfo {author} {\bibfnamefont {L.}~\bibnamefont
  {Asteria}}, \bibinfo {author} {\bibfnamefont {D.~T.}\ \bibnamefont {Tran}},
  \bibinfo {author} {\bibfnamefont {T.}~\bibnamefont {Ozawa}}, \bibinfo
  {author} {\bibfnamefont {M.}~\bibnamefont {Tarnowski}}, \bibinfo {author}
  {\bibfnamefont {B.~S.}\ \bibnamefont {Rem}}, \bibinfo {author} {\bibfnamefont
  {N.}~\bibnamefont {Fl{\"a}schner}}, \bibinfo {author} {\bibfnamefont
  {K.}~\bibnamefont {Sengstock}}, \bibinfo {author} {\bibfnamefont
  {N.}~\bibnamefont {Goldman}}, \ and\ \bibinfo {author} {\bibfnamefont
  {C.}~\bibnamefont {Weitenberg}},\ }\href@noop {} {\bibfield  {journal}
  {\bibinfo  {journal} {Nature Physics}\ }\textbf {\bibinfo {volume} {49}},\
  \bibinfo {pages} {1} (\bibinfo {year} {2019})}\BibitemShut {NoStop}%
\bibitem [{\citenamefont {Coh}\ \emph {et~al.}(2011)\citenamefont {Coh},
  \citenamefont {Vanderbilt}, \citenamefont {Malashevich},\ and\ \citenamefont
  {Souza}}]{Coh:2011gq}%
  \BibitemOpen
  \bibfield  {author} {\bibinfo {author} {\bibfnamefont {S.}~\bibnamefont
  {Coh}}, \bibinfo {author} {\bibfnamefont {D.}~\bibnamefont {Vanderbilt}},
  \bibinfo {author} {\bibfnamefont {A.}~\bibnamefont {Malashevich}}, \ and\
  \bibinfo {author} {\bibfnamefont {I.}~\bibnamefont {Souza}},\ }\href@noop {}
  {\bibfield  {journal} {\bibinfo  {journal} {Physical Review B}\ }\textbf
  {\bibinfo {volume} {83}},\ \bibinfo {pages} {085108} (\bibinfo {year}
  {2011})}\BibitemShut {NoStop}%
\bibitem [{\citenamefont {Essin}\ \emph {et~al.}(2009)\citenamefont {Essin},
  \citenamefont {Moore},\ and\ \citenamefont {Vanderbilt}}]{Essin:2009kb}%
  \BibitemOpen
  \bibfield  {author} {\bibinfo {author} {\bibfnamefont {A.~M.}\ \bibnamefont
  {Essin}}, \bibinfo {author} {\bibfnamefont {J.~E.}\ \bibnamefont {Moore}}, \
  and\ \bibinfo {author} {\bibfnamefont {D.}~\bibnamefont {Vanderbilt}},\
  }\href@noop {} {\bibfield  {journal} {\bibinfo  {journal} {Physical Review
  Letters}\ }\textbf {\bibinfo {volume} {102}},\ \bibinfo {pages} {146805}
  (\bibinfo {year} {2009})}\BibitemShut {NoStop}%
\bibitem [{\citenamefont {Qi}\ \emph {et~al.}(2008)\citenamefont {Qi},
  \citenamefont {Hughes},\ and\ \citenamefont {Zhang}}]{QiHughesZhang08}%
  \BibitemOpen
  \bibfield  {author} {\bibinfo {author} {\bibfnamefont {X.-L.}\ \bibnamefont
  {Qi}}, \bibinfo {author} {\bibfnamefont {T.~L.}\ \bibnamefont {Hughes}}, \
  and\ \bibinfo {author} {\bibfnamefont {S.-C.}\ \bibnamefont {Zhang}},\ }\href
  {\doibase 10.1103/PhysRevB.78.195424} {\bibfield  {journal} {\bibinfo
  {journal} {Phys. Rev. B}\ }\textbf {\bibinfo {volume} {78}},\ \bibinfo
  {pages} {195424} (\bibinfo {year} {2008})}\BibitemShut {NoStop}%
\bibitem [{\citenamefont {Rauch}\ \emph {et~al.}(2018)\citenamefont {Rauch},
  \citenamefont {Olsen}, \citenamefont {Vanderbilt},\ and\ \citenamefont
  {Souza}}]{rauch2018geometric}%
  \BibitemOpen
  \bibfield  {author} {\bibinfo {author} {\bibfnamefont {T.}~\bibnamefont
  {Rauch}}, \bibinfo {author} {\bibfnamefont {T.}~\bibnamefont {Olsen}},
  \bibinfo {author} {\bibfnamefont {D.}~\bibnamefont {Vanderbilt}}, \ and\
  \bibinfo {author} {\bibfnamefont {I.}~\bibnamefont {Souza}},\ }\href@noop {}
  {\bibfield  {journal} {\bibinfo  {journal} {Physical Review B}\ }\textbf
  {\bibinfo {volume} {98}},\ \bibinfo {pages} {115108} (\bibinfo {year}
  {2018})}\BibitemShut {NoStop}%
\bibitem [{\citenamefont {Bianco}\ and\ \citenamefont
  {Resta}(2011)}]{bianco2011mapping}%
  \BibitemOpen
  \bibfield  {author} {\bibinfo {author} {\bibfnamefont {R.}~\bibnamefont
  {Bianco}}\ and\ \bibinfo {author} {\bibfnamefont {R.}~\bibnamefont {Resta}},\
  }\href@noop {} {\bibfield  {journal} {\bibinfo  {journal} {Physical Review
  B}\ }\textbf {\bibinfo {volume} {84}},\ \bibinfo {pages} {241106} (\bibinfo
  {year} {2011})}\BibitemShut {NoStop}%
\bibitem [{Sup()}]{SupplementalHOTI}%
  \BibitemOpen
  \href@noop {} {\bibinfo  {journal} {See Supplemental Material below for
  further details on spectral properties of chiral HOTIs, details of the
  calculation of circular dichroism, and a deeper analysis of the local Chern
  marker in topological materials breaking time-reversal symmetry}\
  }\BibitemShut {NoStop}%
\bibitem [{\citenamefont {Tran}\ \emph {et~al.}(2015)\citenamefont {Tran},
  \citenamefont {Dauphin}, \citenamefont {Goldman},\ and\ \citenamefont
  {Gaspard}}]{tran2015topological}%
  \BibitemOpen
\bibfield  {journal} {  }\bibfield  {author} {\bibinfo {author} {\bibfnamefont
  {D.-T.}\ \bibnamefont {Tran}}, \bibinfo {author} {\bibfnamefont
  {A.}~\bibnamefont {Dauphin}}, \bibinfo {author} {\bibfnamefont
  {N.}~\bibnamefont {Goldman}}, \ and\ \bibinfo {author} {\bibfnamefont
  {P.}~\bibnamefont {Gaspard}},\ }\href@noop {} {\bibfield  {journal} {\bibinfo
   {journal} {Physical Review B}\ }\textbf {\bibinfo {volume} {91}},\ \bibinfo
  {pages} {085125} (\bibinfo {year} {2015})}\BibitemShut {NoStop}%
\bibitem [{\citenamefont {Groth}\ \emph {et~al.}(2014)\citenamefont {Groth},
  \citenamefont {Wimmer}, \citenamefont {Akhmerov},\ and\ \citenamefont
  {Waintal}}]{Groth_2014}%
  \BibitemOpen
  \bibfield  {author} {\bibinfo {author} {\bibfnamefont {C.~W.}\ \bibnamefont
  {Groth}}, \bibinfo {author} {\bibfnamefont {M.}~\bibnamefont {Wimmer}},
  \bibinfo {author} {\bibfnamefont {A.~R.}\ \bibnamefont {Akhmerov}}, \ and\
  \bibinfo {author} {\bibfnamefont {X.}~\bibnamefont {Waintal}},\ }\href
  {\doibase 10.1088/1367-2630/16/6/063065} {\bibfield  {journal} {\bibinfo
  {journal} {New Journal of Physics}\ }\textbf {\bibinfo {volume} {16}},\
  \bibinfo {pages} {063065} (\bibinfo {year} {2014})}\BibitemShut {NoStop}%
\bibitem [{\citenamefont {Mong}\ \emph {et~al.}(2010)\citenamefont {Mong},
  \citenamefont {Essin},\ and\ \citenamefont {Moore}}]{Mong:2010ka}%
  \BibitemOpen
  \bibfield  {author} {\bibinfo {author} {\bibfnamefont {R.~S.~K.}\
  \bibnamefont {Mong}}, \bibinfo {author} {\bibfnamefont {A.~M.}\ \bibnamefont
  {Essin}}, \ and\ \bibinfo {author} {\bibfnamefont {J.~E.}\ \bibnamefont
  {Moore}},\ }\href@noop {} {\bibfield  {journal} {\bibinfo  {journal}
  {Physical Review B}\ }\textbf {\bibinfo {volume} {81}},\ \bibinfo {pages}
  {245209} (\bibinfo {year} {2010})}\BibitemShut {NoStop}%
\bibitem [{\citenamefont {Varjas}\ \emph
  {et~al.}(2019{\natexlab{a}})\citenamefont {Varjas}, \citenamefont {Fruchart},
  \citenamefont {Akhmerov},\ and\ \citenamefont
  {Perez-Piskunow}}]{Varjas:2019wz}%
  \BibitemOpen
  \bibfield  {author} {\bibinfo {author} {\bibfnamefont {D.}~\bibnamefont
  {Varjas}}, \bibinfo {author} {\bibfnamefont {M.}~\bibnamefont {Fruchart}},
  \bibinfo {author} {\bibfnamefont {A.~R.}\ \bibnamefont {Akhmerov}}, \ and\
  \bibinfo {author} {\bibfnamefont {P.}~\bibnamefont {Perez-Piskunow}},\
  }\href@noop {} {\bibfield  {journal} {\bibinfo  {journal} {arXiv.org}\ }
  (\bibinfo {year} {2019}{\natexlab{a}})},\ \Eprint
  {http://arxiv.org/abs/1905.02215v1} {1905.02215v1} \BibitemShut {NoStop}%
\bibitem [{\citenamefont {Wei{\ss}e}\ \emph
  {et~al.}(2006{\natexlab{a}})\citenamefont {Wei{\ss}e}, \citenamefont
  {Wellein}, \citenamefont {Alvermann},\ and\ \citenamefont
  {Fehske}}]{Weisse:2006goa}%
  \BibitemOpen
  \bibfield  {author} {\bibinfo {author} {\bibfnamefont {A.}~\bibnamefont
  {Wei{\ss}e}}, \bibinfo {author} {\bibfnamefont {G.}~\bibnamefont {Wellein}},
  \bibinfo {author} {\bibfnamefont {A.}~\bibnamefont {Alvermann}}, \ and\
  \bibinfo {author} {\bibfnamefont {H.}~\bibnamefont {Fehske}},\ }\href@noop {}
  {\bibfield  {journal} {\bibinfo  {journal} {Reviews of Modern Physics}\
  }\textbf {\bibinfo {volume} {78}},\ \bibinfo {pages} {275} (\bibinfo {year}
  {2006}{\natexlab{a}})}\BibitemShut {NoStop}%
\bibitem [{\citenamefont {Xu}\ \emph {et~al.}(2019)\citenamefont {Xu},
  \citenamefont {Song}, \citenamefont {Wang}, \citenamefont {Weng},\ and\
  \citenamefont {Dai}}]{Xu:2019ui}%
  \BibitemOpen
  \bibfield  {author} {\bibinfo {author} {\bibfnamefont {Y.}~\bibnamefont
  {Xu}}, \bibinfo {author} {\bibfnamefont {Z.}~\bibnamefont {Song}}, \bibinfo
  {author} {\bibfnamefont {Z.}~\bibnamefont {Wang}}, \bibinfo {author}
  {\bibfnamefont {H.}~\bibnamefont {Weng}}, \ and\ \bibinfo {author}
  {\bibfnamefont {X.}~\bibnamefont {Dai}},\ }\href {\doibase
  10.1103/PhysRevLett.122.256402} {\bibfield  {journal} {\bibinfo  {journal}
  {Phys. Rev. Lett.}\ }\textbf {\bibinfo {volume} {122}},\ \bibinfo {pages}
  {256402} (\bibinfo {year} {2019})}\BibitemShut {NoStop}%
\bibitem [{\citenamefont {Hou}\ and\ \citenamefont {Wu}(2019)}]{Hou:2019wd}%
  \BibitemOpen
  \bibfield  {author} {\bibinfo {author} {\bibfnamefont {Y.~S.}\ \bibnamefont
  {Hou}}\ and\ \bibinfo {author} {\bibfnamefont {R.~Q.}\ \bibnamefont {Wu}},\
  }\href {http://arxiv.org/abs/1902.03372v1} {\bibfield  {journal} {\bibinfo
  {journal} {arXiv.org}\ } (\bibinfo {year} {2019})},\ \Eprint
  {http://arxiv.org/abs/1902.03372v1} {1902.03372v1} \BibitemShut {NoStop}%
\bibitem [{\citenamefont {Zhang}\ \emph {et~al.}(2019)\citenamefont {Zhang},
  \citenamefont {Shi}, \citenamefont {Zhu}, \citenamefont {Xing}, \citenamefont
  {Zhang},\ and\ \citenamefont {Wang}}]{Zhang:2019fk}%
  \BibitemOpen
  \bibfield  {author} {\bibinfo {author} {\bibfnamefont {D.}~\bibnamefont
  {Zhang}}, \bibinfo {author} {\bibfnamefont {M.}~\bibnamefont {Shi}}, \bibinfo
  {author} {\bibfnamefont {T.}~\bibnamefont {Zhu}}, \bibinfo {author}
  {\bibfnamefont {D.}~\bibnamefont {Xing}}, \bibinfo {author} {\bibfnamefont
  {H.}~\bibnamefont {Zhang}}, \ and\ \bibinfo {author} {\bibfnamefont
  {J.}~\bibnamefont {Wang}},\ }\href {\doibase 10.1103/PhysRevLett.122.206401}
  {\bibfield  {journal} {\bibinfo  {journal} {Physical Review Letters}\
  }\textbf {\bibinfo {volume} {122}},\ \bibinfo {pages} {206401} (\bibinfo
  {year} {2019})}\BibitemShut {NoStop}%
\bibitem [{\citenamefont {Otrokov}\ \emph {et~al.}(2018)\citenamefont
  {Otrokov}, \citenamefont {Klimovskikh}, \citenamefont {Bentmann},
  \citenamefont {Zeugner}, \citenamefont {Aliev}, \citenamefont {Gass},
  \citenamefont {Wolter}, \citenamefont {Koroleva}, \citenamefont {Estyunin},
  \citenamefont {Shikin}, \citenamefont {Blanco-Rey}, \citenamefont {Hoffmann},
  \citenamefont {Vyazovskaya}, \citenamefont {Eremeev}, \citenamefont
  {Koroteev}, \citenamefont {Amiraslanov}, \citenamefont {Babanly},
  \citenamefont {Mamedov}, \citenamefont {Abdullayev}, \citenamefont {Zverev},
  \citenamefont {B{\"u}chner}, \citenamefont {Schwier}, \citenamefont {Kumar},
  \citenamefont {Kimura}, \citenamefont {Petaccia}, \citenamefont {Di~Santo},
  \citenamefont {Vidal}, \citenamefont {Schatz}, \citenamefont {Ki{\ss}ner},
  \citenamefont {Min}, \citenamefont {Moser}, \citenamefont {Peixoto},
  \citenamefont {Reinert}, \citenamefont {Ernst}, \citenamefont {Echenique},
  \citenamefont {Isaeva},\ and\ \citenamefont {Chulkov}}]{Otrokov:2018vx}%
  \BibitemOpen
  \bibfield  {author} {\bibinfo {author} {\bibfnamefont {M.~M.}\ \bibnamefont
  {Otrokov}}, \bibinfo {author} {\bibfnamefont {I.~I.}\ \bibnamefont
  {Klimovskikh}}, \bibinfo {author} {\bibfnamefont {H.}~\bibnamefont
  {Bentmann}}, \bibinfo {author} {\bibfnamefont {A.}~\bibnamefont {Zeugner}},
  \bibinfo {author} {\bibfnamefont {Z.~S.}\ \bibnamefont {Aliev}}, \bibinfo
  {author} {\bibfnamefont {S.}~\bibnamefont {Gass}}, \bibinfo {author}
  {\bibfnamefont {A.~U.~B.}\ \bibnamefont {Wolter}}, \bibinfo {author}
  {\bibfnamefont {A.~V.}\ \bibnamefont {Koroleva}}, \bibinfo {author}
  {\bibfnamefont {D.}~\bibnamefont {Estyunin}}, \bibinfo {author}
  {\bibfnamefont {A.~M.}\ \bibnamefont {Shikin}}, \bibinfo {author}
  {\bibfnamefont {M.}~\bibnamefont {Blanco-Rey}}, \bibinfo {author}
  {\bibfnamefont {M.}~\bibnamefont {Hoffmann}}, \bibinfo {author}
  {\bibfnamefont {A.~Y.}\ \bibnamefont {Vyazovskaya}}, \bibinfo {author}
  {\bibfnamefont {S.~V.}\ \bibnamefont {Eremeev}}, \bibinfo {author}
  {\bibfnamefont {Y.~M.}\ \bibnamefont {Koroteev}}, \bibinfo {author}
  {\bibfnamefont {I.~R.}\ \bibnamefont {Amiraslanov}}, \bibinfo {author}
  {\bibfnamefont {M.~B.}\ \bibnamefont {Babanly}}, \bibinfo {author}
  {\bibfnamefont {N.~T.}\ \bibnamefont {Mamedov}}, \bibinfo {author}
  {\bibfnamefont {N.~A.}\ \bibnamefont {Abdullayev}}, \bibinfo {author}
  {\bibfnamefont {V.~N.}\ \bibnamefont {Zverev}}, \bibinfo {author}
  {\bibfnamefont {B.}~\bibnamefont {B{\"u}chner}}, \bibinfo {author}
  {\bibfnamefont {E.~F.}\ \bibnamefont {Schwier}}, \bibinfo {author}
  {\bibfnamefont {S.}~\bibnamefont {Kumar}}, \bibinfo {author} {\bibfnamefont
  {A.}~\bibnamefont {Kimura}}, \bibinfo {author} {\bibfnamefont
  {L.}~\bibnamefont {Petaccia}}, \bibinfo {author} {\bibfnamefont
  {G.}~\bibnamefont {Di~Santo}}, \bibinfo {author} {\bibfnamefont {R.~C.}\
  \bibnamefont {Vidal}}, \bibinfo {author} {\bibfnamefont {S.}~\bibnamefont
  {Schatz}}, \bibinfo {author} {\bibfnamefont {K.}~\bibnamefont {Ki{\ss}ner}},
  \bibinfo {author} {\bibfnamefont {C.-H.}\ \bibnamefont {Min}}, \bibinfo
  {author} {\bibfnamefont {S.~K.}\ \bibnamefont {Moser}}, \bibinfo {author}
  {\bibfnamefont {T.~R.~F.}\ \bibnamefont {Peixoto}}, \bibinfo {author}
  {\bibfnamefont {F.}~\bibnamefont {Reinert}}, \bibinfo {author} {\bibfnamefont
  {A.}~\bibnamefont {Ernst}}, \bibinfo {author} {\bibfnamefont {P.~M.}\
  \bibnamefont {Echenique}}, \bibinfo {author} {\bibfnamefont {A.}~\bibnamefont
  {Isaeva}}, \ and\ \bibinfo {author} {\bibfnamefont {E.~V.}\ \bibnamefont
  {Chulkov}},\ }\href {http://arxiv.org/abs/1809.07389v1} {\bibfield  {journal}
  {\bibinfo  {journal} {arXiv.org}\ } (\bibinfo {year} {2018})},\ \Eprint
  {http://arxiv.org/abs/1809.07389v1} {1809.07389v1} \BibitemShut {NoStop}%
\bibitem [{\citenamefont {Gong}\ \emph {et~al.}(2018)\citenamefont {Gong},
  \citenamefont {Guo}, \citenamefont {Li}, \citenamefont {Zhu}, \citenamefont
  {Liao}, \citenamefont {Liu}, \citenamefont {Zhang}, \citenamefont {Gu},
  \citenamefont {Tang}, \citenamefont {Feng}, \citenamefont {Zhang},
  \citenamefont {Li}, \citenamefont {Song}, \citenamefont {Wang}, \citenamefont
  {Yu}, \citenamefont {Chen}, \citenamefont {Wang}, \citenamefont {Yao},
  \citenamefont {Duan}, \citenamefont {Xu}, \citenamefont {Zhang},
  \citenamefont {Ma}, \citenamefont {Xue},\ and\ \citenamefont
  {He}}]{Gong:2018to}%
  \BibitemOpen
  \bibfield  {author} {\bibinfo {author} {\bibfnamefont {Y.}~\bibnamefont
  {Gong}}, \bibinfo {author} {\bibfnamefont {J.}~\bibnamefont {Guo}}, \bibinfo
  {author} {\bibfnamefont {J.}~\bibnamefont {Li}}, \bibinfo {author}
  {\bibfnamefont {K.}~\bibnamefont {Zhu}}, \bibinfo {author} {\bibfnamefont
  {M.}~\bibnamefont {Liao}}, \bibinfo {author} {\bibfnamefont {X.}~\bibnamefont
  {Liu}}, \bibinfo {author} {\bibfnamefont {Q.}~\bibnamefont {Zhang}}, \bibinfo
  {author} {\bibfnamefont {L.}~\bibnamefont {Gu}}, \bibinfo {author}
  {\bibfnamefont {L.}~\bibnamefont {Tang}}, \bibinfo {author} {\bibfnamefont
  {X.}~\bibnamefont {Feng}}, \bibinfo {author} {\bibfnamefont {D.}~\bibnamefont
  {Zhang}}, \bibinfo {author} {\bibfnamefont {W.}~\bibnamefont {Li}}, \bibinfo
  {author} {\bibfnamefont {C.}~\bibnamefont {Song}}, \bibinfo {author}
  {\bibfnamefont {L.}~\bibnamefont {Wang}}, \bibinfo {author} {\bibfnamefont
  {P.}~\bibnamefont {Yu}}, \bibinfo {author} {\bibfnamefont {X.}~\bibnamefont
  {Chen}}, \bibinfo {author} {\bibfnamefont {Y.}~\bibnamefont {Wang}}, \bibinfo
  {author} {\bibfnamefont {H.}~\bibnamefont {Yao}}, \bibinfo {author}
  {\bibfnamefont {W.}~\bibnamefont {Duan}}, \bibinfo {author} {\bibfnamefont
  {Y.}~\bibnamefont {Xu}}, \bibinfo {author} {\bibfnamefont {S.-C.}\
  \bibnamefont {Zhang}}, \bibinfo {author} {\bibfnamefont {X.}~\bibnamefont
  {Ma}}, \bibinfo {author} {\bibfnamefont {Q.-K.}\ \bibnamefont {Xue}}, \ and\
  \bibinfo {author} {\bibfnamefont {K.}~\bibnamefont {He}},\ }\href
  {http://arxiv.org/abs/1809.07926v1} {\bibfield  {journal} {\bibinfo
  {journal} {arXiv.org}\ } (\bibinfo {year} {2018})},\ \Eprint
  {http://arxiv.org/abs/1809.07926v1} {1809.07926v1} \BibitemShut {NoStop}%
\bibitem [{\citenamefont {{Tiwari}}\ \emph {et~al.}(2019)\citenamefont
  {{Tiwari}}, \citenamefont {{Li}}, \citenamefont {{Bernevig}}, \citenamefont
  {{Neupert}},\ and\ \citenamefont {{Parameswaran}}}]{Tiwari2019}%
  \BibitemOpen
  \bibfield  {author} {\bibinfo {author} {\bibfnamefont {A.}~\bibnamefont
  {{Tiwari}}}, \bibinfo {author} {\bibfnamefont {M.-H.}\ \bibnamefont {{Li}}},
  \bibinfo {author} {\bibfnamefont {B.~A.}\ \bibnamefont {{Bernevig}}},
  \bibinfo {author} {\bibfnamefont {T.}~\bibnamefont {{Neupert}}}, \ and\
  \bibinfo {author} {\bibfnamefont {S.~A.}\ \bibnamefont {{Parameswaran}}},\
  }\href@noop {} {\bibfield  {journal} {\bibinfo  {journal} {arXiv e-prints}\
  ,\ \bibinfo {eid} {arXiv:1905.11421}} (\bibinfo {year} {2019})},\ \Eprint
  {http://arxiv.org/abs/1905.11421} {arXiv:1905.11421 [cond-mat.str-el]}
  \BibitemShut {NoStop}%
\bibitem [{\citenamefont {Bianco}\ and\ \citenamefont
  {Resta}(2013)}]{BiancoResta2013}%
  \BibitemOpen
  \bibfield  {author} {\bibinfo {author} {\bibfnamefont {R.}~\bibnamefont
  {Bianco}}\ and\ \bibinfo {author} {\bibfnamefont {R.}~\bibnamefont {Resta}},\
  }\href {\doibase 10.1103/PhysRevLett.110.087202} {\bibfield  {journal}
  {\bibinfo  {journal} {Phys. Rev. Lett.}\ }\textbf {\bibinfo {volume} {110}},\
  \bibinfo {pages} {087202} (\bibinfo {year} {2013})}\BibitemShut {NoStop}%
\bibitem [{\citenamefont {Caio}\ \emph {et~al.}(2019)\citenamefont {Caio},
  \citenamefont {M{\"o}ller}, \citenamefont {Cooper},\ and\ \citenamefont
  {Bhaseen}}]{Caio:2019jk}%
  \BibitemOpen
  \bibfield  {author} {\bibinfo {author} {\bibfnamefont {M.~D.}\ \bibnamefont
  {Caio}}, \bibinfo {author} {\bibfnamefont {G.}~\bibnamefont {M{\"o}ller}},
  \bibinfo {author} {\bibfnamefont {N.~R.}\ \bibnamefont {Cooper}}, \ and\
  \bibinfo {author} {\bibfnamefont {M.~J.}\ \bibnamefont {Bhaseen}},\ }\href
  {\doibase 10.1038/s41567-018-0390-7} {\bibfield  {journal} {\bibinfo
  {journal} {Nature Physics}\ }\textbf {\bibinfo {volume} {15}},\ \bibinfo
  {pages} {257} (\bibinfo {year} {2019})}\BibitemShut {NoStop}%
\bibitem [{\citenamefont {Ardila}\ \emph {et~al.}(2018)\citenamefont {Ardila},
  \citenamefont {Heyl},\ and\ \citenamefont {Eckardt}}]{ardila2018measuring}%
  \BibitemOpen
  \bibfield  {author} {\bibinfo {author} {\bibfnamefont {L.~A.~P.}\
  \bibnamefont {Ardila}}, \bibinfo {author} {\bibfnamefont {M.}~\bibnamefont
  {Heyl}}, \ and\ \bibinfo {author} {\bibfnamefont {A.}~\bibnamefont
  {Eckardt}},\ }\href@noop {} {\bibfield  {journal} {\bibinfo  {journal}
  {Physical review letters}\ }\textbf {\bibinfo {volume} {121}},\ \bibinfo
  {pages} {260401} (\bibinfo {year} {2018})}\BibitemShut {NoStop}%
\bibitem [{\citenamefont {Mitchell}\ \emph {et~al.}(2018)\citenamefont
  {Mitchell}, \citenamefont {Nash}, \citenamefont {Hexner}, \citenamefont
  {Turner},\ and\ \citenamefont {Irvine}}]{mitchell2018amorphous}%
  \BibitemOpen
  \bibfield  {author} {\bibinfo {author} {\bibfnamefont {N.~P.}\ \bibnamefont
  {Mitchell}}, \bibinfo {author} {\bibfnamefont {L.~M.}\ \bibnamefont {Nash}},
  \bibinfo {author} {\bibfnamefont {D.}~\bibnamefont {Hexner}}, \bibinfo
  {author} {\bibfnamefont {A.~M.}\ \bibnamefont {Turner}}, \ and\ \bibinfo
  {author} {\bibfnamefont {W.~T.}\ \bibnamefont {Irvine}},\ }\href@noop {}
  {\bibfield  {journal} {\bibinfo  {journal} {Nature Physics}\ ,\ \bibinfo
  {pages} {1}} (\bibinfo {year} {2018})}\BibitemShut {NoStop}%
\bibitem [{\citenamefont {Schine}\ \emph {et~al.}(2019)\citenamefont {Schine},
  \citenamefont {Chalupnik}, \citenamefont {Can}, \citenamefont {Gromov},\ and\
  \citenamefont {Simon}}]{schine2019electromagnetic}%
  \BibitemOpen
  \bibfield  {author} {\bibinfo {author} {\bibfnamefont {N.}~\bibnamefont
  {Schine}}, \bibinfo {author} {\bibfnamefont {M.}~\bibnamefont {Chalupnik}},
  \bibinfo {author} {\bibfnamefont {T.}~\bibnamefont {Can}}, \bibinfo {author}
  {\bibfnamefont {A.}~\bibnamefont {Gromov}}, \ and\ \bibinfo {author}
  {\bibfnamefont {J.}~\bibnamefont {Simon}},\ }\href@noop {} {\bibfield
  {journal} {\bibinfo  {journal} {Nature}\ }\textbf {\bibinfo {volume} {565}},\
  \bibinfo {pages} {173} (\bibinfo {year} {2019})}\BibitemShut {NoStop}%
\bibitem [{\citenamefont {Okada}\ \emph {et~al.}(2016)\citenamefont {Okada},
  \citenamefont {Takahashi}, \citenamefont {Mogi}, \citenamefont {Yoshimi},
  \citenamefont {Tsukazaki}, \citenamefont {Takahashi}, \citenamefont {Ogawa},
  \citenamefont {Kawasaki},\ and\ \citenamefont {Tokura}}]{Okada:2016tk}%
  \BibitemOpen
  \bibfield  {author} {\bibinfo {author} {\bibfnamefont {K.~N.}\ \bibnamefont
  {Okada}}, \bibinfo {author} {\bibfnamefont {Y.}~\bibnamefont {Takahashi}},
  \bibinfo {author} {\bibfnamefont {M.}~\bibnamefont {Mogi}}, \bibinfo {author}
  {\bibfnamefont {R.}~\bibnamefont {Yoshimi}}, \bibinfo {author} {\bibfnamefont
  {A.}~\bibnamefont {Tsukazaki}}, \bibinfo {author} {\bibfnamefont {K.~S.}\
  \bibnamefont {Takahashi}}, \bibinfo {author} {\bibfnamefont {N.}~\bibnamefont
  {Ogawa}}, \bibinfo {author} {\bibfnamefont {M.}~\bibnamefont {Kawasaki}}, \
  and\ \bibinfo {author} {\bibfnamefont {Y.}~\bibnamefont {Tokura}},\ }\href
  {https://doi.org/10.1038/ncomms12245} {\bibfield  {journal} {\bibinfo
  {journal} {Nature Communications}\ }\textbf {\bibinfo {volume} {7}},\
  \bibinfo {pages} {12245} (\bibinfo {year} {2016})}\BibitemShut {NoStop}%
\bibitem [{\citenamefont {Olsen}\ \emph {et~al.}(2017)\citenamefont {Olsen},
  \citenamefont {Taherinejad}, \citenamefont {Vanderbilt},\ and\ \citenamefont
  {Souza}}]{Olsen:2017bz}%
  \BibitemOpen
  \bibfield  {author} {\bibinfo {author} {\bibfnamefont {T.}~\bibnamefont
  {Olsen}}, \bibinfo {author} {\bibfnamefont {M.}~\bibnamefont {Taherinejad}},
  \bibinfo {author} {\bibfnamefont {D.}~\bibnamefont {Vanderbilt}}, \ and\
  \bibinfo {author} {\bibfnamefont {I.}~\bibnamefont {Souza}},\ }\href@noop {}
  {\bibfield  {journal} {\bibinfo  {journal} {Physical Review B}\ }\textbf
  {\bibinfo {volume} {95}},\ \bibinfo {pages} {075137} (\bibinfo {year}
  {2017})}\BibitemShut {NoStop}%
\bibitem [{\citenamefont {Malashevich}\ \emph {et~al.}(2010)\citenamefont
  {Malashevich}, \citenamefont {Souza}, \citenamefont {Coh},\ and\
  \citenamefont {Vanderbilt}}]{Malashevich:2010hn}%
  \BibitemOpen
  \bibfield  {author} {\bibinfo {author} {\bibfnamefont {A.}~\bibnamefont
  {Malashevich}}, \bibinfo {author} {\bibfnamefont {I.}~\bibnamefont {Souza}},
  \bibinfo {author} {\bibfnamefont {S.}~\bibnamefont {Coh}}, \ and\ \bibinfo
  {author} {\bibfnamefont {D.}~\bibnamefont {Vanderbilt}},\ }\href@noop {}
  {\bibfield  {journal} {\bibinfo  {journal} {New Journal of Physics}\ }\textbf
  {\bibinfo {volume} {12}},\ \bibinfo {pages} {053032} (\bibinfo {year}
  {2010})}\BibitemShut {NoStop}%
\bibitem [{\citenamefont {Taherinejad}\ and\ \citenamefont
  {Vanderbilt}(2015)}]{Taherinejad2015}%
  \BibitemOpen
  \bibfield  {author} {\bibinfo {author} {\bibfnamefont {M.}~\bibnamefont
  {Taherinejad}}\ and\ \bibinfo {author} {\bibfnamefont {D.}~\bibnamefont
  {Vanderbilt}},\ }\href {\doibase 10.1103/PhysRevLett.114.096401} {\bibfield
  {journal} {\bibinfo  {journal} {Phys. Rev. Lett.}\ }\textbf {\bibinfo
  {volume} {114}},\ \bibinfo {pages} {096401} (\bibinfo {year}
  {2015})}\BibitemShut {NoStop}%
\bibitem [{\citenamefont {PythTB}()}]{PythTB}%
  \BibitemOpen
  \bibfield  {author} {\bibinfo {author} {\bibnamefont {PythTB}},\ }\href
  {http://www.physics.rutgers.edu/pythtb.} {\bibinfo  {journal}
  {http://www.physics.rutgers.edu/pythtb}\ }\BibitemShut {NoStop}%
\bibitem [{\citenamefont {Varjas}\ \emph
  {et~al.}(2019{\natexlab{b}})\citenamefont {Varjas}, \citenamefont {Lau},
  \citenamefont {P{\"o}yh{\"o}nen}, \citenamefont {Akhmerov}, \citenamefont
  {Pikulin},\ and\ \citenamefont {Fulga}}]{Varjas:2019tp}%
  \BibitemOpen
\bibfield  {journal} {  }\bibfield  {author} {\bibinfo {author} {\bibfnamefont
  {D.}~\bibnamefont {Varjas}}, \bibinfo {author} {\bibfnamefont
  {A.}~\bibnamefont {Lau}}, \bibinfo {author} {\bibfnamefont {K.}~\bibnamefont
  {P{\"o}yh{\"o}nen}}, \bibinfo {author} {\bibfnamefont {A.~R.}\ \bibnamefont
  {Akhmerov}}, \bibinfo {author} {\bibfnamefont {D.~I.}\ \bibnamefont
  {Pikulin}}, \ and\ \bibinfo {author} {\bibfnamefont {I.~C.}\ \bibnamefont
  {Fulga}},\ }\href@noop {} {\bibfield  {journal} {\bibinfo  {journal}
  {arXiv.org}\ } (\bibinfo {year} {2019}{\natexlab{b}})},\ \Eprint
  {http://arxiv.org/abs/1904.07242v2} {1904.07242v2} \BibitemShut {NoStop}%
\bibitem [{\citenamefont {Resta}(2006)}]{Resta2006}%
  \BibitemOpen
  \bibfield  {author} {\bibinfo {author} {\bibfnamefont {R.}~\bibnamefont
  {Resta}},\ }\href {\doibase 10.1063/1.2176604} {\bibfield  {journal}
  {\bibinfo  {journal} {The Journal of Chemical Physics}\ }\textbf {\bibinfo
  {volume} {124}},\ \bibinfo {pages} {104104} (\bibinfo {year}
  {2006})}\BibitemShut {NoStop}%
\bibitem [{\citenamefont {Resta}(2011)}]{Resta2011}%
  \BibitemOpen
  \bibfield  {author} {\bibinfo {author} {\bibfnamefont {R.}~\bibnamefont
  {Resta}},\ }\href {\doibase 10.1140/epjb/e2010-10874-4} {\bibfield  {journal}
  {\bibinfo  {journal} {The European Physical Journal B}\ }\textbf {\bibinfo
  {volume} {79}},\ \bibinfo {pages} {121} (\bibinfo {year} {2011})}\BibitemShut
  {NoStop}%
\bibitem [{\citenamefont {Irsigler}\ \emph {et~al.}(2019)\citenamefont
  {Irsigler}, \citenamefont {Zheng},\ and\ \citenamefont
  {Hofstetter}}]{irsigler2019microscopic}%
  \BibitemOpen
  \bibfield  {author} {\bibinfo {author} {\bibfnamefont {B.}~\bibnamefont
  {Irsigler}}, \bibinfo {author} {\bibfnamefont {J.-H.}\ \bibnamefont {Zheng}},
  \ and\ \bibinfo {author} {\bibfnamefont {W.}~\bibnamefont {Hofstetter}},\
  }\href@noop {} {\bibfield  {journal} {\bibinfo  {journal} {arXiv preprint
  arXiv:1904.03091}\ } (\bibinfo {year} {2019})}\BibitemShut {NoStop}%
\bibitem [{\citenamefont {Bianco}(2014)}]{Biancothesis}%
  \BibitemOpen
  \bibfield  {author} {\bibinfo {author} {\bibfnamefont {R.}~\bibnamefont
  {Bianco}},\ }\href {http://hdl.handle.net/10077/9959} {\bibfield  {journal}
  {\bibinfo  {journal} {PhD Thesis}\ } (\bibinfo {year} {2014})}\BibitemShut
  {NoStop}%
\bibitem [{\citenamefont {Wei{\ss}e}\ \emph
  {et~al.}(2006{\natexlab{b}})\citenamefont {Wei{\ss}e}, \citenamefont
  {Wellein}, \citenamefont {Alvermann},\ and\ \citenamefont
  {Fehske}}]{weisse2006kernel}%
  \BibitemOpen
  \bibfield  {author} {\bibinfo {author} {\bibfnamefont {A.}~\bibnamefont
  {Wei{\ss}e}}, \bibinfo {author} {\bibfnamefont {G.}~\bibnamefont {Wellein}},
  \bibinfo {author} {\bibfnamefont {A.}~\bibnamefont {Alvermann}}, \ and\
  \bibinfo {author} {\bibfnamefont {H.}~\bibnamefont {Fehske}},\ }\href@noop {}
  {\bibfield  {journal} {\bibinfo  {journal} {Reviews of modern physics}\
  }\textbf {\bibinfo {volume} {78}},\ \bibinfo {pages} {275} (\bibinfo {year}
  {2006}{\natexlab{b}})}\BibitemShut {NoStop}%
\end{thebibliography}%

\section*{Supplemental Material}

\section{A. Spectral properties of chiral HOTIs}
\begin{figure}
    \centering
    \includegraphics[scale=1.0]{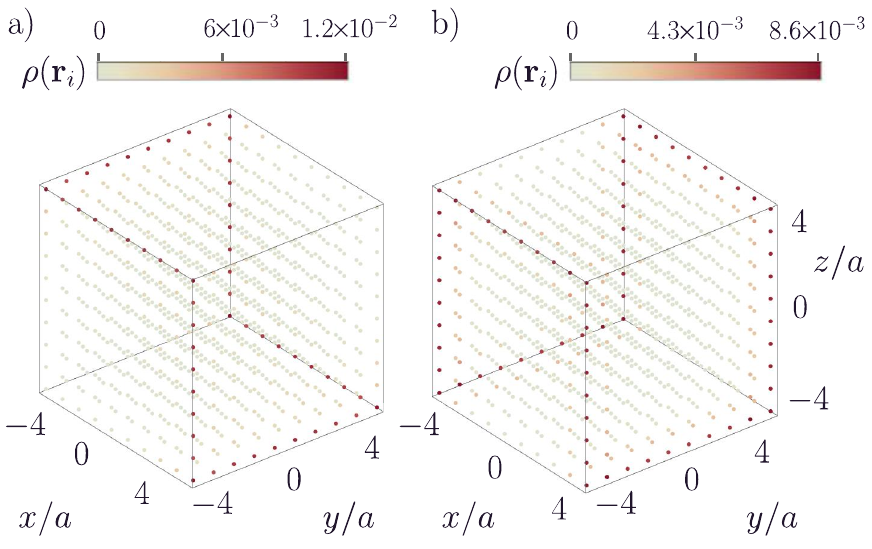}
    \caption{Local electronic density of (a) the $\mathcal{I}$ symmetric HOTI and (b) the $C_4^{z}\mathcal{I}$ symmetric HOTI defined in Eq.~\eqref{eq:model}
    of the main text.}
    \label{fig:HOTIs_DoS}
\end{figure}

The chiral HOTIs we study in the main text, and also Chern insulators, present one dimensional chiral metallic states at their hinges and edges. These states are fingerprints of their non--trivial topology. Through exact diagonalization, one can solve the tight-binding model (Eq.~\eqref{eq:model}
in the main text) and compute the eigenstates $\{ \ket{n} \}$ in the Wannier basis $\{ \ket{\textbf{r}_{i}} \}$. These metallic states are identified with the eigenstates energetically well--located inside the insulating gap, and the local density
\begin{equation}
  \rho_{n}(\textbf{r}_{i}) = |\braket{\textbf{r}_{i}|n}|^{2}  \  ,
\end{equation} 
can be used to check their dimensionality and position. The hinge states of the $\mathcal{I}$- and $C_{4}^{z}\mathcal{I}$ symmetric HOTIs studied in the main text are shown in Fig. \ref{fig:HOTIs_DoS}.

The slab geometry is achieved assuming periodic boundary conditions in two of the three spatial directions, leading to three different slab geometries. There are no hinges in the slab geometry, therefore the energy spectrum is fully gapped. The slab gap $E_g$ used in the main text is taken to be the minimum gap between the three possible slab geometries.

\section{\label{DetailsCD}B. Further details on the computation of the circular dichroism}

\subsection{Computation of the absorption rate}
 
In the main text we presented the expression of the absorption rate, $\Gamma_{\pm}(\omega)$ (Eq.~\eqref{eq:CDEnergy}), to analyze how circular dichroism is quantized in chiral HOTIs. Here we explain the necessary steps to determine such expression. 

Consider a chiral HOTI subjected to an optical signal at frequency $\omega$ polarized in the $xy$  plane. Then the full Hamiltonian at time $t$ is given by
\begin{equation}
    \hat{H}_{\pm} = \hat{H}_{0} + 2E\left[ \cos(\omega t)\hat{x}\pm \sin(\omega t)\hat{y} \right]  \  ,
\end{equation}
where $\hat{H}_{0}$ is the HOTI Hamiltonian defined in Eq.~\eqref{eq:model} 
of the main text, $E$ is the amplitude of the electric field, $\pm$ indicates left- or right handed polarization and $\hat{x}, \hat{y}$ are position operators. Part of this light is absorbed by the electrons of the system, bringing them from the initial ground state $\ket{\Psi_{0}}$ to an excited state $\ket{\Psi(t)}$. The probability of excitation is

\begin{equation}
    P_{\pm}(t) = 1 - | \braket{\Psi(t) | \Psi_{0}} |^{2}  \  ,
\end{equation}
If the electric field amplitude is small enough, the wave function $\ket{\Psi(t)}$ can be computed using time-dependent perturbation theory up to first order. The resulting absorption rate $\Gamma_{\pm}(\omega)=P_{\pm}(t)/t$ is Eq. (2)
%\eqref{eq:CDEnergy} 
in the main text. For long enough times this absorption rate is given by Fermi's Golden Rule
\begin{IEEEeqnarray}{l}
\nonumber
    \dfrac{\Gamma_{\pm}(\omega)}{2\pi E^{2}} =
    \sum_{n,m} \big| \bra{m}\left( \hat{x}\pm i\hat{y} \right) \ket{n} \big|^{2} \delta(E_{m}-E_{n}-\hbar\omega),\\
\end{IEEEeqnarray}
where $\ket{n}$ ($\ket{m}$) are occupied (unoccupied) states and $E_{n}$ ($E_{m}$) their energies. The total circular dichroism is obtained by taking the difference $(\Gamma_{+}-\Gamma_{-})/2$ and integrating over all frequencies as defined in the main text. This difference can be expressed as
\begin{equation}
    \Delta\Gamma(\infty) = -2i\pi E^{2} \text{Tr} \Big\{ \big[ \hat{Q}\hat{x},\hat{P}\hat{y} \big] \Big\} \equiv E^2 \sum_{\text{all } \textbf{r}_{i}}C_{xy}(\textbf{r}_{i})  \  ,
    \label{eq:suppC}
\end{equation}
where $\hat{P}$ $(\hat{Q})$ is the projector over occupied (unoccupied) states. 
Independently of dimensionality, the total absorption rate is proportional to the total Hall conductivity $\sigma_{\alpha\beta}$ of the sample~\cite{Tran2017} where $\alpha\beta$ is the polarization plane of the optical signal.\\

\begin{figure}
    \centering
    \includegraphics[scale=0.14]{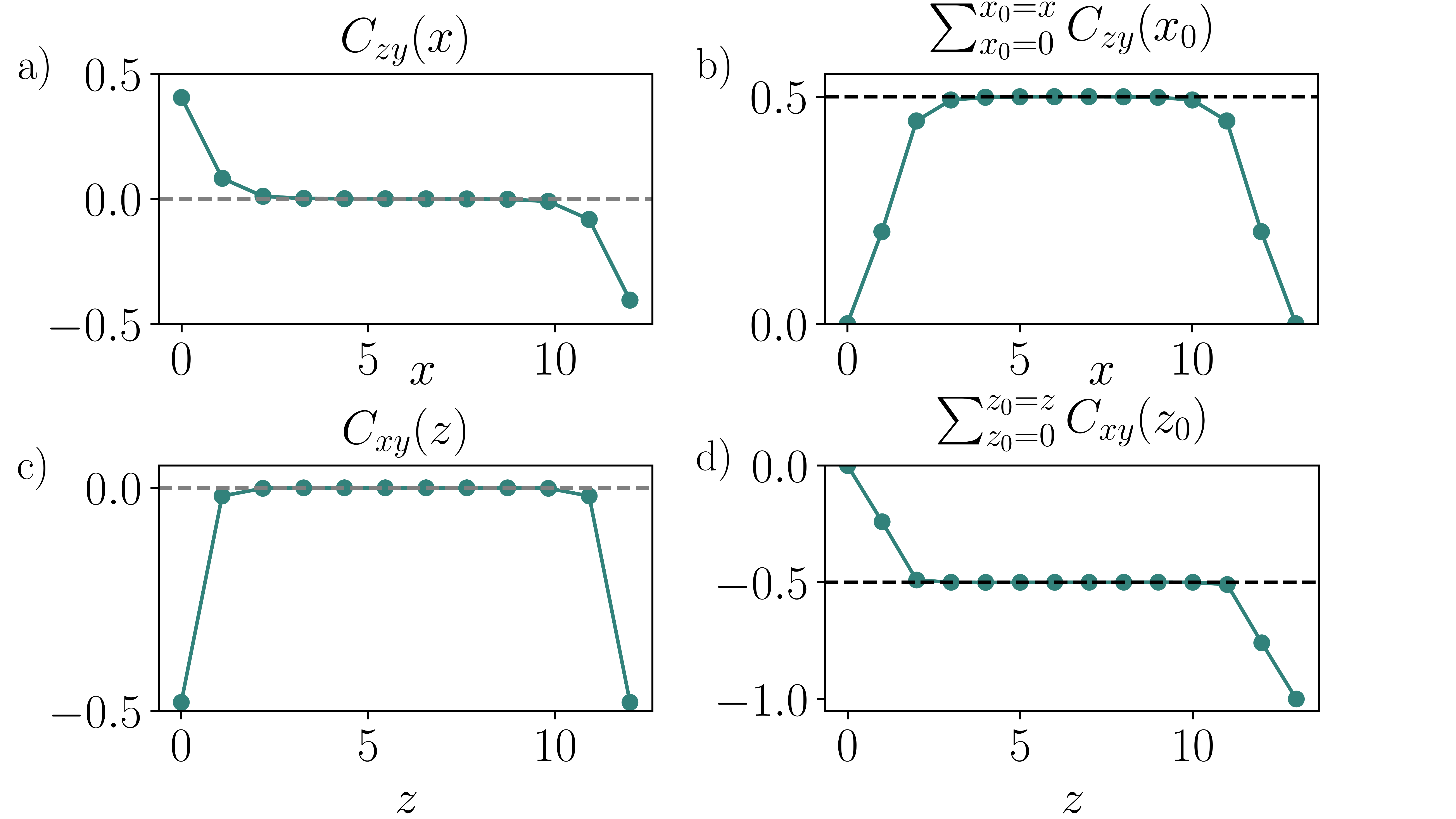}
    \caption{
    Layer-resolved [(a),(c)] and layer-summed [(b),(d)] Chern number of the $C_{4}^{z}\mathcal{I}$-symmetric HOTI in the $zy$ [(a),(b)] and $xy$ [(c),(d)] polarization planes, for a finite slab with periodic boundary conditions in the polarization plane. The Chern number summed over all layers is proportional to the circular dichroic signal $\Delta\Gamma_{b\to b}$.  The curves are calculated following~\cite{varnava2018surfaces} using the open source code \texttt{PythTB}~\cite{PythTB} and a slab thickness of $N=12$.}
    \label{fig:AHC}
\end{figure}

\subsection{Layer-resolved Chern number}
In the case of periodic boundary conditions in the polarization plane~$\alpha\beta$, let us resolve the right hand side of Eq.~\eqref{eq:suppC} along the direction $\gamma$ perpendicular to the polarization plane. We show this layer-resolved Chern number
$C_{\alpha\beta}(\gamma)$ in Fig.~\ref{fig:AHC} (a) and (c) for the $C^z_4\mathcal{I}$-symmetric HOTI. In the infinite slab geometry, the Hall conductivity can be calculated following~\cite{varnava2018surfaces}, by summing the layer-resolved Chern number over the direction of propagation $\gamma$. Fig.~\ref{fig:AHC} (a) and (c) show that the only sizable contributions are due to the surface states which contribute $\pm1/2$, and extend as much as two layers from the surface. This justifies our choice in Fig.~\ref{fig:tetrahedral} (a) in the main text where we chose the two last layers as our surface. Summing over the whole slab we find that the contributions of opposite surfaces either cancel, leading to $C=0$ [see Fig.~\ref{fig:AHC} (b)], or add up to a quantized value $C=-1$ [see Fig.~\ref{fig:AHC} (d)].

The results in Fig.~\ref{fig:AHC}, and using Eq.~\eqref{eq:suppC} show that for periodic boundary conditions, where only bulk-bulk transitions are present, the circular dichroism is quantized to 
\begin{equation}
    \Delta\Gamma_{b\rightarrow b} = CE^{2}A  \ .
\end{equation}
as we use in the main text.

\begin{figure}
    \includegraphics[scale=1.0]{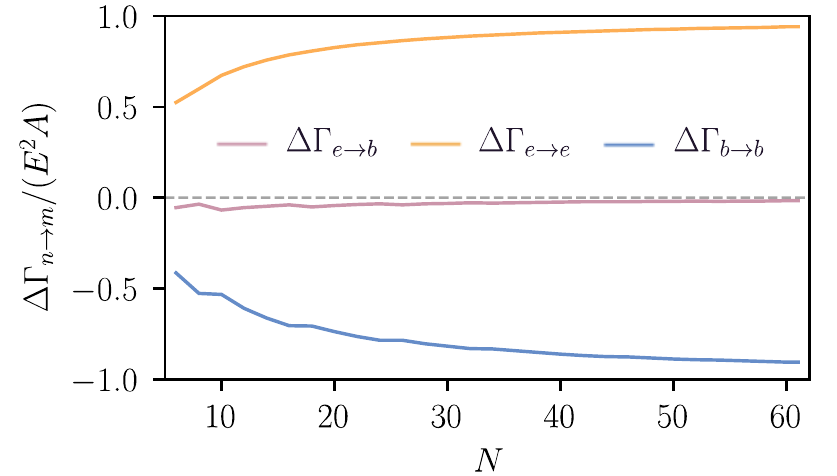}
    \caption{Circular dichroic signal of the 2D Chern insulator Eq.~\eqref{eq:appCI} with $N_\mathrm{tot}=N\times N$ sites computed using exact diagonalization.
    Edge-bulk and bulk-edge transitions give a negligible contribution, while we see the convergence of the bulk-bulk and edge-edge contributions towards their quantized thermodynamic limit value.}
    \label{fig:2DCI_Convergence}
\end{figure}

\subsection{Edge-bulk and bulk-edge contributions in Chern insulators}

The quantization of bulk responses is only ensured in the thermodynamic limit. It is in such case that bulk states of a finite system are independent of boundary conditions and resemble those of periodic boundary conditions. As a consequence, our finite size results do not show an exact quantization of circular dichroism but rather a trend as we approach the thermodynamic limit. Due to the 2D character of Chern insulators, it is possible to calculate the absorption rate of larger systems with exact diagonalization~\cite{Varjas:2019wz,Varjas:2019tp}. This allows us to show the quantization of edge-bulk, bulk-bulk and edge-edge contributions in the thermodynamic limit to larger precision (see Fig.~\ref{fig:2DCI_Convergence}). The Chern insulator model we have considered is
\begin{equation}
\label{eq:appCI}
    H = \left( M - t\sum_{i=1}^{2}\cos\left( k_{i}a \right) \right) \sigma_{3} + \lambda\sum_{i=1}^{2}\sin(k_{i}a)\sigma_{i}  \ ,
\end{equation}
where $M/t=3/2$ and $\lambda/t=-1/2$.

\subsection{Effect of a finite chemical potential}
In the main text we have assumed half-filling, and thus the chemical potential $\mu$ sits exactly at the center of the gap. This assumption made the edge-bulk and bulk-edge contributions equivalent in Fig.~\ref{fig:CD} b)
of the main text. In general these two types of transitions can be different.
However, our results remain unchanged so long as the chemical potential remains in the gap. Out of all possible transitions, the bulk-bulk transitions are still quantized if the chemical potential is within the gap. The edge-bulk and bulk-edge contributions vanish, independent of the chemical potential. Therefore, the sum of all edge-edge transitions must exactly compensate the sum of bulk-bulk transitions with an equal and opposite Chern number. Summing over all edge-edge transitions implies integrating $\Delta\Gamma(E_g)$, as for the case where $\mu=0$.

\section{C.  Further details on the computation of the Local Chern marker}

In Fig.~\ref{fig:tetrahedral}
of the main text, we see that $C_{xy}(\textbf{r}_{i})$ reaches its thermodynamic value much faster in the bulk than in the hinges. This is related to the contrasting scaling of correlation functions in these two regions: states are respectively exponentially and algebraically localized in the bulk and the hinges. As a result, at a given bulk point, the main contribution to the local Chern marker is determined by the nearest points \cite{Resta2006,Resta2011,irsigler2019microscopic}, while it involves larger non-local contributions for the hinges. A more formal way to understand this is to consider the expression of the local Chern marker in terms of the density correlator
\begin{equation}
    \rho_{n}(\textbf{r}_{i},\textbf{r}_{j}) = \braket{\textbf{r}_{i}|n}\braket{n|\textbf{r}_{j}}  \ ,
\end{equation}
which reads~\cite{irsigler2019microscopic}
\begin{widetext}
\begin{equation}
    C_{xy}(\textbf{r}_{i}) = -4\pi\sum_{\textbf{r}_{j},\textbf{r}_{k}} \sum_{\substack{E_{l}<0 \\ E_{m}>0 \\ E_{n}<0 }}\text{Im}\left[ \rho_{l}(\textbf{r}_{i},\textbf{r}_{j})\rho_{m}(\textbf{r}_{j},\textbf{r}_{k})\rho_{n}(\textbf{r}_{k},\textbf{r}_{i}) \right] x_{j}y_{k} \  .
\end{equation}
\end{widetext}
Since the magnitude of $\rho_{n}(\textbf{r}_{i},\textbf{r}_{j})$ is proportional to the modulus of the density of the state $\ket{n}$ at $\textbf{r}_{i}$ and at $\textbf{r}_{j}$, bulk-edge and edge-bulk contributions are suppressed here as in the case of circular dichroism. Because of the long-range density correlations within the hinge, the local Chern marker at a hinge point $\textbf{r}_{i}$ receives significant contributions from other hinge points $\left( \textbf{r}_{j},\textbf{r}_{k} \right)$, even though they are not spatially close. 

\subsection{Convergence of the hinge contribution}

As shown in Fig.~\ref{fig:tetrahedral} (a) and (b) in the main text at a point $\textbf{r}_{i}$ in the top and bottom surface of the $C_4^z\mathcal{I}$-symmetric HOTI $C_{xy}(\textbf{r}_{i})$ is quantized to $1/2$, similar to how a bulk point in a Chern insulator defined by Eq.~\eqref{eq:appCI} is quantized to $\pm1$~\cite{Biancothesis}. 
In contrast, the value of the local Chern marker at a hinge point need not be quantized. We analyze the non-negligible local Chern marker contributions, which come from the surface bulk ($b$), the horizontal ($h$) and the vertical ($v$) hinges (see Fig.~\ref{fig:tetrahedral}
of the main text). Note that, due to the $C_{4}^{z}\mathcal{I}$-symmetry, the contribution of all vertices is the same and we can democratically associate one vertex to each hinge.
In an finite system, these contributions verify the following sum rule
\begin{equation}
\label{eq:appsum}
  0 = \sum_{\textbf{r}_{i}} C_{xy}( \textbf{r}_{i} ) = 2C_{b}+4C_{h}+4C_{v}  \   .
\end{equation}
$C_b$ is quantized to $\pm1/2 N_s$ where $N_s$ is the number of surface area points defined in blue in Fig.~\ref{fig:tetrahedral} (a) of the main text (in the thermodynamic limit $N_s \simeq N^2$). The sum $C_h + C_v$ is thus quantized (in units of $N_s$), but $C_{h}$ and $C_{v}$ are not forced by $C_{4}^{z}\mathcal{I}$ symmetry to be quantized independently. We note this is unlike the Chern insulator Eq.~\eqref{eq:appCI}, which is invariant under $C_{4}^{z}$, implying that the Chern marker at each edge is forced to be $C_{\mathrm{edge}}=-C/4$. 

Finally, for Fig.~\ref{fig:tetrahedral} (d) we computed the local Chern marker using the efficient numerical approximation based on the Kernel Polynomial Method \cite{weisse2006kernel} implemented via the \texttt{Kwant} Python package~\cite{Groth_2014}. It allows us to reach larger system sizes than the ones achievable with exact diagonalization. 
The projector $\hat{P}$ is approximated as a polynomial in powers of $\hat{H}$ of degree inversely proportional to the broadening parameter indicated in the main text (see~\cite{Varjas:2019wz} for details).
We have checked convergence by observing that a broadening of $10^{-3}$ reproduces the local Chern marker results from exact diagonalization for small $N$, and that our results change by less than $0.1\% $ when the broadening is reduced by a factor of $10$.

\end{document}